\documentclass[12pt,preprint]{aastex}

\slugcomment{The L1615/L1616 PMS population: \today}

\shorttitle{The star formation in the L1615/L1616 cometary cloud}
\shortauthors{Gandolfi et al.}

\begin{document}


\title{The star formation in the L1615/L1616 Cometary Cloud}


\author{
Davide~Gandolfi\altaffilmark{1,2,3},
Juan~M.~Alcal\'a\altaffilmark{2},
Silvio~Leccia\altaffilmark{2},
Antonio~Frasca\altaffilmark{1},
Loredana~Spezzi\altaffilmark{1,2},
Elvira~Covino\altaffilmark{2},
Leonardo~Testi\altaffilmark{4},
Ettore~Marilli\altaffilmark{1},
Jouni~Kainulainen\altaffilmark{5}
}

%%\affil{}


\altaffiltext{1}{INAF, Osservatorio Astrofisico di Catania, Via S. Sofia 78, 95123 Catania, Italy; dgandolfi@oact.inaf.it}
\altaffiltext{2}{INAF, Osservatorio Astronomico di Capodimonte, Salita Moiariello 16, 80131  Napoli, Italy; jmae@na.astro.it}
\altaffiltext{3}{Th\"uringer Landessternwarte Tautenburg, Sternwarte 5, 07778 Tautenburg, Germany; davide@tls-tautenburg.de}
\altaffiltext{4}{INAF, Osservatorio Astronomico di Arcetri, L.go E. Fermi 5, 50125 Firenze, Italy; lt@arcetri.astro.it}
\altaffiltext{5}{Observatory, PO Box 14, FIN-00014 University of Helsinki, Finland; jouni.kainulainen@helsinki.fi}


\begin{abstract}

The present work aims at performing a comprehensive census and
characterisation of the pre-main sequence (PMS) population in
the cometary cloud L1615/L1616, in order to assess the significance
of the triggered star formation scenario and investigate the impact
of massive stars on its star formation history and mass spectrum.
Our study is based on $UBVR_CI_C$ and $JHKs$ photometry, as well as
optical multi-object spectroscopy. We performed a physical
parametrisation of the young stellar population in L1615/L1616.
We identified 25 new T~Tauri stars mainly projected on the dense head
of the cometary cloud, almost doubling the current number of known members.
We studied the spatial distribution of the cloud members as a function
of the age and H$\alpha$ emission. The star formation efficiency in
the cloud is $\sim$~7--8\,\%, as expected for molecular clouds in the vicinity
of OB associations. The slope of the initial mass function (IMF), in
the mass range $0.1{\le}M{\le}5.5$~$M_{\odot}$, is consistent with that of
other T and OB associations, providing further support of an universal
IMF down to the hydrogen burning limit, regardless of environmental
conditions. The cometary appearance, as well as the high star formation
efficiency, can be explained in terms of triggered star formation induced
by the strong UV radiation from OB stars or supernovae shockwaves. The
age spread as well as both the spatial and age distribution of the PMS
objects provide strong evidence of sequential, multiple events and
possibly still ongoing star formation activity in the cloud.

\end{abstract}





\section{Introduction}
\label{sec:Intro}

The Lynds clouds L1615 and L1616 \citep{Lynds62} are both located
at an angular distance of about $6\degr$ West of the Orion OB1
associations. These clouds actually form a cometary-shaped single
cloud with a ``head-tail'' distribution subtending about $40\arcmin$
\citep[5.2~pc assuming a distance of 450~pc;][]{Alcala04} roughly
in the East-West direction. The dense head of the cloud complex
(i.e. L1616 only), pointing toward East and facing the bright
Orion-belt stars, harbours the IRAS Small Scale Structure
X0504-034 \citep{Helou88,Ramesh95} and a bright reflection nebula, 
NGC\,1788 \citep[= DG\,51, Ced\,40, vdB\,33, RNO\,35, LBN\,916;][]{Lynds65}, 
which is illuminated by a small cluster of stars \citep{Stanke02}. The two
intermediate-mass stars HD\,293815 and Kiso\,A-0974~15 are the
brightest visible members of this cluster.

L1615/L1616, like many other cometary clouds off the main Orion
star forming regions, clearly shows evidence of ongoing star formation
activity which might have been triggered by the strong impact of
the UV radiation from the massive, luminous stars in the Orion
complex \citep{Maddalena86,Stanke02, Alcala04,Kun04,Lee05,Lee07}.
The illumination of dense clumps in molecular clouds by OB stars
could be responsible for their collapse and subsequent star formation.
The UV radiation from the OB stars may sweep the molecular material
of the cloud into a cometary shape with a dense core located at
the head of the cometary cloud.

The radiation and wind of OB stars may also have an important
impact on the mass accretion during the star formation process.
While in a T~association a protostar may accumulate a significant
fraction of mass, the mass accretion of a low-mass protostar in a
region exposed to the wind of OB stars can be terminated earlier
because of the photo-evaporation of the circumstellar matter
\citep{Kroupa01,Kroupa02}. Therefore, many low-mass protostars may
not complete their accretion and hence can result as brown dwarfs
(BDs). This mechanism might affect the low-mass end of the initial
mass function (IMF). Recent studies have provided some information 
about the shape of the IMF in the very low-mass and sub-stellar regimes.
While the IMF in the Orion Nebula Cluster appears to rise below
0.1 M$_{\odot}$ \citep{Hillenbrand00}, in T~association like
Taurus-Auriga and Cha\,II there is some indication of a deficit of
sub-stellar objects \citep{Luhman00,Briceno02,Spezzi08}. Other studies
of the young cluster IC\,348, which is devoid of very massive stars,
have also revealed a deficit of BDs relative to the Orion Nebula
Cluster \citep{Preibisch03,Muench03,Lada06}.

Because of its vicinity to the Orion OB associations the L1615/L1616
cometary cloud constitutes an ideal laboratory to investigate the
triggered star formation scenario.
The most recent census of the pre-main sequence (PMS) stars in L1616
was provided by \citet{Alcala04}, who presented a multi-wavelength
study of the region, from X-ray to near-infrared wavelengths. They
found 22 new low-mass PMS stars distributed mainly to the East of
L1616, in about 1~square degree field. By adding the 22~new PMS stars
to the previously confirmed members of the cloud
\citep{Cohen79,Sterzik95,Nakano95,Stanke02} and counting the
millimeter radio-source, i.e. the Class-0 protostar MMS1A found by
\citet{Stanke02}, \citet{Alcala04} ended up with a sample of 33~young
stellar objects associated with L1616. However, the latter work could
not investigate important aspects of the star formation in this cloud,
like the IMF and the impact of environmental conditions on the mass
spectrum, because their sample was rather incomplete.

Therefore, the aim of the present study is to perform a comprehensive
census and characterisation of the PMS population in L1615/L1616,
in order to investigate the star formation history, the relevance
of the triggered scenario, and to study the IMF.
To this aim, we report both optical and near-infrared
observations, as well as multi-object optical spectroscopy in
L1615/L1616. An important goal of our study is to compare the mass
function in L1615/L1616 with that of other T and OB~associations.

% This will allow us to further investigate whether the environmental
% star-forming conditions could have an impact on the IMF in the very-low
% mass domain.

The outline of the paper is as follows: we describe our observations
and data reduction in Section~\ref{sec:PhotObs} and \ref{sec:Spec-Obs};
the results are reported in Section~\ref{sec:PMS-Ident}, while the data
analysis and the physical proprieties of the new PMS stars are presented
in Section~\ref{sec:Data analysis}. Our discussion and conclusions are
developed in Section~\ref{sec:Discussion}. Some details on the
spectral-type classification as well as reddening, radius, and
luminosity determination are given in the Appendix~\ref{App-A} and
\ref{App-B}, respectively.

\section{Photometry}
\label{sec:PhotObs}

\subsection{Optical}
\label{sec:OptPhot}

%% In a manner similar to \objectname authors can provide links to dataset
%% hosted at participating data centers via the \dataset{} command.  The
%% second curly bracket argument is printed in the text while the first
%% parentheses argument serves as the valid data set identifier.  Large
%% lists of data set are best provided in a table (see Table 3 for an example).
%% Valid data set identifiers should be obtained from the data center that
%% is currently hosting the data.

Most of the optical photometry comes from previous $B$, $V$,
$R_C$ and $I_C$ broad-band imaging observations performed
by \citet{Alcala04}, who obtained CCD mosaic images with the
Wide Field Imager (WFI) camera at the ESO/MPIA 2.2\,m telescope
(La Silla Observatory, Chile, program No. 64.I-0355), covering
a sky-area of about $36{\arcmin}\times34{\arcmin}$ around
NGC\,1788. The pre-reduction of the raw images, as well as the
astrometric and photometric calibration have been already discussed
in \citet{Alcala04}. In order to improve the photometry of the
blended sources, we carried out point spread function fitting
photometry on the images using the DAOPHOT package \citep{Stetson87}
under the IRAF\footnote{IRAF is distributed by the National Optical
Astronomy Observatory, which is operated by the Association of the
Universities for Research in Astronomy, inc. (AURA) under
cooperative agreement with the National Science Foundation.}
environment.

%%%%%%%%Serra La Nave Photometric Observations%%%%%%%%%%%%%%%%%%%%%
Further photometric observations of six previously known members of
L1615/L1616 \citep{Alcala04}, still lacking optical photometry and
falling outside the sky-area covered by WFI, were carried out in
November 17th and 18th, 2006 with the 91\,cm telescope at Catania
Astrophysical Observatory (OAC). The observations were performed in
the Johnson $UBV$ standard system under photometric sky conditions,
by using a photon-counting single-head photometer equipped with an
EMI\,9893QA/350 photomultiplier. In order to determine the
transformation coefficients to the Johnson standard system, several
standard stars selected from the General Catalogue of Photometric
Data \citep[GCPD,][]{Mermilliod97} and from \citet{Landolt92} were
also observed in between the targets observations. The observing
set-up and the data reduction were the standard ones already adopted during
previous observational campaigns of low-mass stars
\citep[see, e.g.,][]{Marilli07}.

\subsection{Near-infrared}
\label{sec:NIRPhot}

Near-infrared $J$, $H$ and $Ks$ photometry of a
$\sim30{\arcmin}\times30{\arcmin}$ region centered on NGC\,1788 was
obtained using the SOFI camera at the ESO NTT telescope
(program No. 70.C-0629) on December 17th, 2002 and January 16th, 2003. 
36 pointings of the $5{\arcmin}\times5{\arcmin}$ field
of view were required to cover the field in each band. 
Several dithered exposures were obtained for a total integration time per
pointing of 80 seconds in the $J$ band and $180$ seconds in the $H$ and
$Ks$ bands. The data were corrected for electronic crosstalk,
differential flat fielding, and illumination following standard
SOFI procedures under IRAF.

Photometric calibration was obtained observing a set of standard stars
from the list of \citet{Persson98} (Las Campanas Observatory standards). 
Source extraction and aperture photometry were performed using the 
SExtractor software package \citep{Bertin96}; aperture corrections were 
estimated from sources away from crowded regions and applied to all 
extracted sources. The limiting magnitudes at 3$\sigma$ level achieved 
in our survey are 19, 18 and 17.5 mag at $J$, $H$, and $Ks$, respectively.

\section{Spectroscopy}
\label{sec:Spec-Obs}

Based on $I_C$ versus $(R_C-I_C)$ colour-magnitude diagram and
theoretical isochrones from \citet{Baraffe98}, \citet{Alcala04}
selected a sample of about 200 PMS star and BD candidates in L1615/L1616,
by using the spectroscopically confirmed young low-mass stars to define
the PMS locus, as shown in their Figure\,12. In addition
to this sample, 8 X-ray emitting stars detected by the
ROSAT All-Sky Survey (RASS), as well as mid-infrared sources
\citep{Stanke02} and H$\alpha$ emission-line objects  \citep{Nakano95},
were also selected as further PMS star candidates by \citet{Alcala04}. 
We performed the spectroscopic follow-up of about 70\,\% of the candidates 
in this sample, down to $I_C\approx19$~mag, in order to definitively 
assess their nature and single out the young objects. Additionally, 
about 30\,\% of field stars with $I_C\lesssim20.0$~mag was also observed 
in order to detect possible ``veiled" PMS stars which might have escaped 
the selection criterion (see later Figure~\ref{Fig:isoc_Bes}).

\begin{figure}
  \centering
  \resizebox{\hsize}{!}{\includegraphics[draft=false]{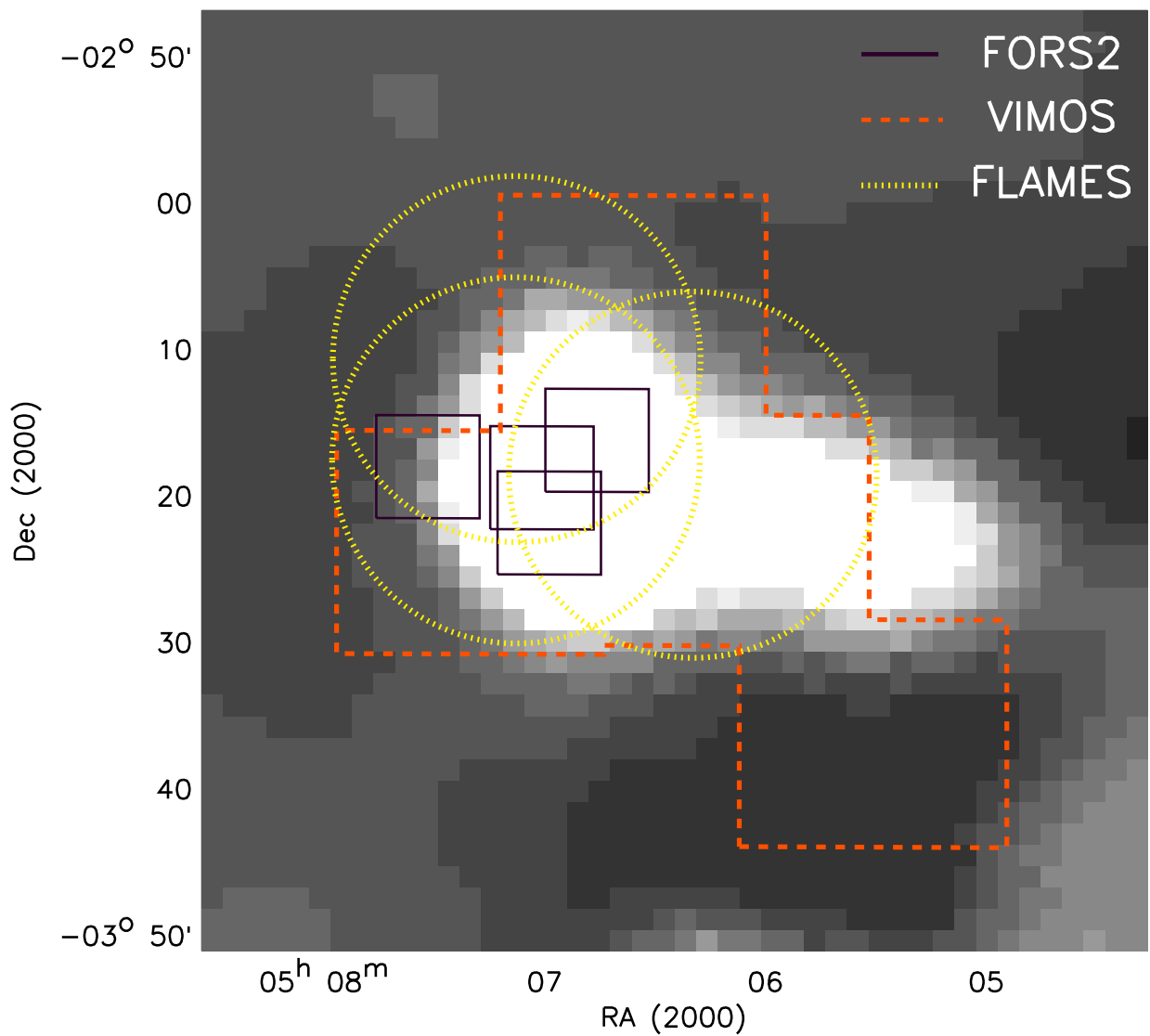}}%[height=6in,width=6in]
  \caption{Sky areas surveyed by the FORS2 (continuous lines), VIMOS
           (dashed lines), and FLAMES (dotted lines) spectroscopic
           follow-ups overlaid on the IRAS $100~\mu$m dust emission
           map covering a sky-area of about $1^\circ\times1^\circ$
           around L1615/L1616. Note that the gaps of about $2\arcmin$
           between the four VIMOS quadrants are not reproduced in
           the figure.}
  \label{Fig:SpecSurvey}
\end{figure}

Intermediate- and low-resolution spectra were acquired during different
runs at the ESO Very Large Telescope (VLT; Paranal Observatory, Chile),
by using the FOcal Reducer and low-dispersion Spectrograph 2
\citep[FORS2;][]{Appenzeller98} and the VIsible Multi-Object
Spectrograph \citep[VIMOS;][]{LeFevre03}. Multi-object
intermediate-resolution spectroscopy was also performed with the Fibre
Large Area Multi-Element Spectrograph \citep[FLAMES;][]{Pasquini02},
but these observations will be described in more detail in a forthcoming
paper. The journal of the FORS2 and VIMOS observations is given in
Table~\ref{Tab:Spec-Log}. The distribution on the sky of the area covered
by the FORS2, VIMOS, and FLAMES spectroscopic follow-ups is shown in
Figure~\ref{Fig:SpecSurvey}. In the following sub-sections a brief
description of the spectroscopic observations and data reduction is
presented.

\clearpage
\clearpage

\subsection{FORS2 observations}
\label{sec:FORS2-Obs}

Intermediate-resolution ($\lambda$/$\Delta\lambda$=$2140$) FORS2
spectra of the PMS candidates mainly projected on the NGC\,1788
reflection nebula were collected in service mode (ESO program
No. 70.C-0536) on the nights 7, 22, 25 February and 8 March 2003,
under mostly clear and stable weather conditions, with seeing
typically in the range $0{\farcs}8-1{\farcs}0$. These observations
were performed setting the spectrograph in the multi-object
spectroscopy mode. A total of 5 mask configurations, each
including 18--19 objects, were observed. Three consecutive spectra
of 960~sec were obtained for each mask configuration in order to
remove cosmic-ray hits and improve the signal to noise (S/N) ratio. Bias,
flat-spectrum lamp, and Ar/He/Ne lamp exposures for each mask were
taken in daytime according to the FORS2 standard calibration plan.
The typical S/N ratio for a $I_C\approx16$~mag star was on the order of 90.

The standard data reduction process was carried out with a
semi-automatic pipeline that we have developed on the basis of both
MIDAS\footnote{MIDAS is developed and maintained by ESO.}
\citep{Warmels91} and IRAF package. The reduction includes bias
subtraction, flat-field division, wavelength calibration, sky
subtraction and one-dimensional spectra extraction. Relative flux
calibration was achieved observing five spectrophotometric standard
stars from \citet{Hamuy92,Hamuy94}.

\subsection{VIMOS observations}
\label{sec:VIMOS-Obs}

The VIMOS observations were performed in service mode (ESO program
No. 074.C-0111) during the period from 9 December 2004 to 12 March
2005. The weather conditions were photometric with seeing varying
between $0{\farcs}6$ and $1{\farcs}2$. The larger sky-area covered
with VIMOS allowed us to extend the spectroscopic survey in a wider
region around the NGC\,1788 reflection nebula\footnote{Note, however,
that the gaps of about $2{\arcmin}$ between the four VIMOS quadrants 
prevented us from performing a uniform coverage.}. Furthermore, we 
carried out both intermediate- ($\lambda$/$\Delta\lambda$=$2500$) and
low-resolution ($\lambda$/$\Delta\lambda$=$580$) spectroscopy, in
order to detect the Li\,{\sc i} $\lambda$6708~{\AA} absorption line
and measure its equivalent width, as well as to obtain a wider
wavelength range for spectral type classification purpose (see
Section~\ref{sec:Spec-Type-Class}).

A total of 3 intermediate- and 6 low-resolution mask configurations,
each including an average number of about 100 targets,
were observed. To efficiently pursue our programme and maximise the
number of observed fields, only one exposure of 900 sec and one of
2000 sec were obtained for each low- and intermediate-resolution
mask configuration, respectively.
The typical S/N ratio at $I_C\approx16$~mag was about 80 and 180 
for the intermediate- and low-resolution spectra, respectively.
%For the low-resolution spectra we found that a single exposure of
%900 sec was a good compromise between reaching an adequate S/N
%ratio and avoiding a high contamination of cosmic ray hits.

Five spectrophotometric standard stars from \citet{Oke90} and
\citet{Hamuy92,Hamuy94} were observed to perform the absolute flux
calibrations. Bias and flat field frames as well as wavelength
calibration exposures were acquired during daytime, following
the service mode VIMOS calibration plan.

The VIMOS spectra were automatically reduced by using the ESO
pipeline, full details of which are given by \citet{Izzo04}. In
order to check the reliability of the data reduction, the spectra
were also independently processed by using the VIMOS Interactive
Pipeline Graphical Interface \citep[VIPGI;][]{Scodeggio05},
obtaining consistent results.

\subsection{FLAMES-GIRAFFE observations}
\label{sec:FLAMES-Obs}

The FLAMES-GIRAFFE observations were performed in visitor mode
(ESO program No. 076.C-0385) during the nights 28 February 2006
and 1, 2 March 2006, under good seeing conditions
($0{\farcs}7-1{\farcs}0$), setting the spectrograph in the MEDUSA
configuration. The low-resolution grating LR06 was used in
conjunction with an order separating filter and a slit width of
$1{\arcsec}$. The adopted configuration yielded a spectral coverage
of about 750~{\AA} (6438--7184~{\AA}) with a mean resolving power
$\lambda$/$\Delta\lambda$=$8600$.

Since these observations are part of a separate study on radial
velocities in L1615/L1616, these data will be described in more
detail in a future paper. Here we have used the FLAMES spectra
of only three objects, namely TTS\,050644.4$-$032913,
RX\,J0507.3$-$0326, and TTS\,050741.0$-$032253
(see Table~\ref{Tab:List-PMS}), for which both FORS2 and VIMOS
observations are missing.

\section{PMS objects in L1615/L1616}
\label{sec:PMS-Ident}

Since lithium is rapidly destroyed in the convective layers of low-mass
stars in the early phases of their stellar evolution \citep{Bodenheimer65},
the presence of strong Li\,{\sc i} $\lambda$6708~{\AA} absorption line
has been considered as the primary criterion for definitely assessing
the PMS nature of the observed stars. The H$\alpha$ emission has been
used as further indicator of youth, though its presence alone
does not guarantee the PMS nature of the observed object.
%In first instance, the presence of H$\alpha$ emission line has not
%been considered by itself as criterion for establishing the PMS
%nature of the objects, but only as further signature confirming
%their youth, especially for classical T Tauri stars (CTTSs).
Therefore, all the K and M stars with H$\alpha$ in emission but lacking
the Li\,{\sc i} absorption line have been rejected and classified as
older dKe and dMe field stars projected onto L1615/L1616.

In addition to the above criteria, the presence of forbidden
emission lines, as well as non-photospheric UV and infrared continuum
excesses (as evidence of bipolar jets, accretion columns, and
circumstellar disk, respectively) have yielded further arguments
on the PMS nature of the young stars identified in L1615/L1616.

\begin{figure}
 \resizebox{\hsize}{!}{\includegraphics[draft=false]{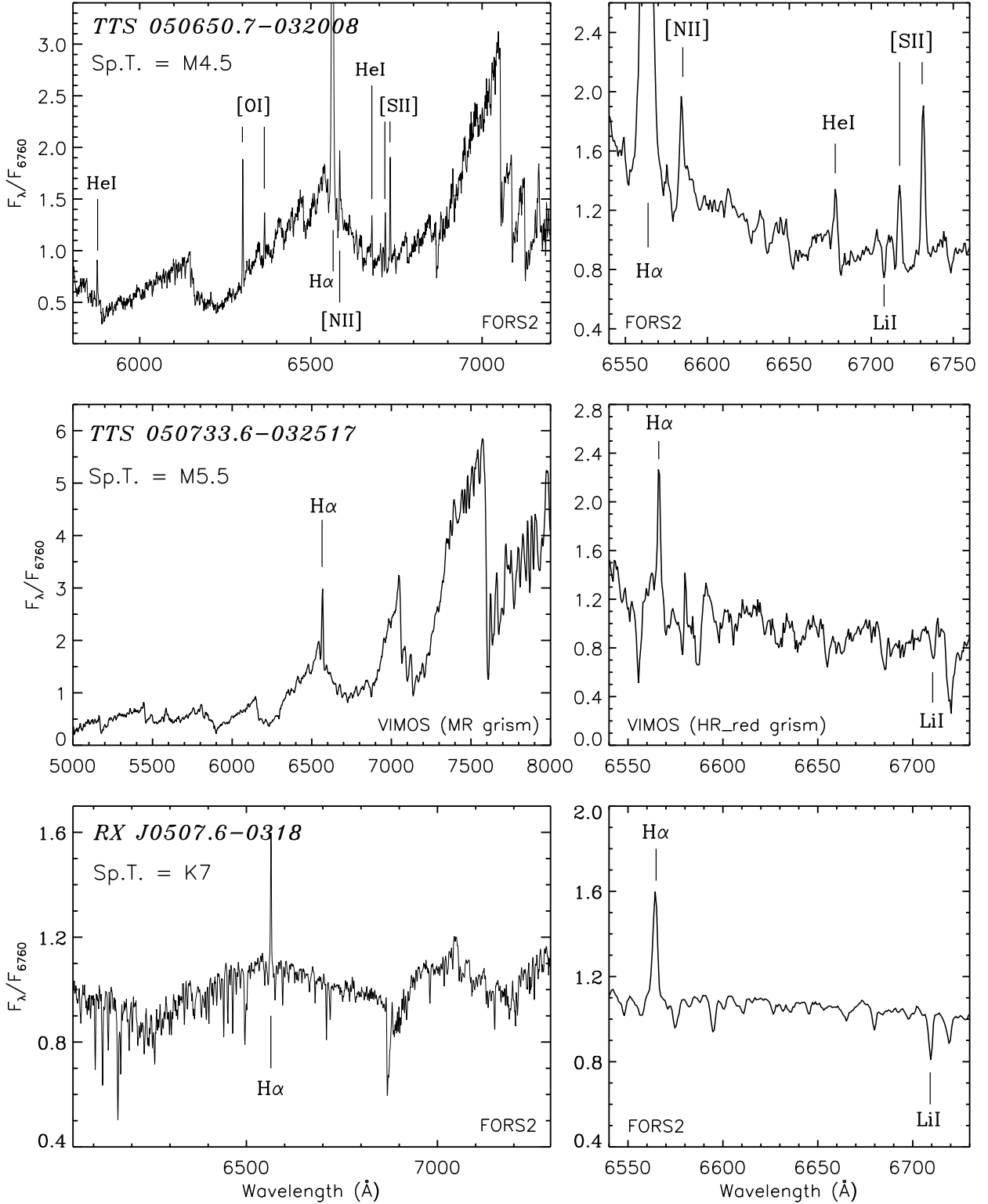}}
 \caption{VIMOS and FORS2 spectra of three PMS objects in L1615/L1616, 
          namely TTS\,050650.7-032008, TTS\,050733.6-032517, and 
          RX\,J0507.6-0318. The spectra have been arbitrarily normalised to 
          the flux at 6760~{\AA}. For each object the whole observed spectrum 
          is shown in the left panel. The spectral range encompassing the 
          H$\alpha$ and Li\,{\sc i} $\lambda$6708~{\AA} lines is plotted in 
          more detail in the right panel.}
 \label{Fig:Spectra}
\end{figure}

Based on these criteria, we have identified 25 new PMS stars with
spectral type later than about K1. By adding these sources to the previously
reported PMS objects \citep{Alcala04}, the number of optically detected
young stars known so far in L1615/L1616 rises to 56 objects\footnote{Note
that the star Kiso\,A-0974~17 \citep{Nakano95,Alcala04} has been
excluded from our list of PMS objects (see below).}.
The full sample is listed in Table~\ref{Tab:List-PMS}; coordinates
are those from the 2\,Micron All Sky Survey \citep[2MASS;][]{Cutri03} 
point source catalogue, which have a precision of about $0{\farcs}1$. 
All the PMS stars revealed by our spectroscopic follow-up,
but not detected in previous X-ray and H$\alpha$ surveys, have been
designated with ``TTS'' followed by their position, according to the
IAU convention. By adding the millimeter Class-0 protostar discovered
by \citet{Stanke02}, which is not reported in Table~\ref{Tab:List-PMS}, the
young population of L1615/L1616 increases to 57. Thus, our spectroscopic
survey almost has doubled the number of confirmed PMS objects associated
to L1615/L1616. Some examples of FORS2 and VIMOS spectra are shown in
Figure~\ref{Fig:Spectra}.
%Figures showing the whole set of spectra are
%included in the electronically available appendix.

In total, 4 out of the 8 X-ray emitting sources listed by
\citet{Alcala04}, namely RX\,J0506.8$-$0305, RX\,J0507.4$-$0317,
RX\,J0507.6$-$0318, and RX\,J0507.3$-$0326, have been spectroscopically
confirmed as PMS stars; the other X-ray sources (RX\,J0506.5$-$0320,\\
RX\,J0506.4$-$0323, RX\,J0506.4$-$0328W, and RX\,J0506.4$-$0328E)
turned out to be field active stars. Note that lithium has been
detected for the first time in the spectrum of the mid-infrared  
source L1616\,MIR4 \citep{Stanke02}.

The S/N ratio of the spectra of the faint sources
TTS\,050654.5$-$032046 and\\
TTS\,050730.9$-$031846 turned out
to be insufficient for a reliable measurement of their Li\,{\sc i}
equivalent width. Nevertheless, the presence of a broad and strong
H$\alpha$ emission in both spectra (Table~\ref{Tab:SpT-W})
can not be explained by chromospheric activity only and is typical
of classical T Tauri stars. Indeed, according to the definition 
by \citet{White03}, the upper limit on the H$\alpha$ equivalent width 
for chromospheric activity, in the spectral range M3--M5.5, is $40$~\AA. 
TTS\,050654.5$-$032046 and TTS\,050730.9$-$031846 have spectral type 
M4 and M5.5 and equivalent width $60$~\AA~ and $290$~\AA, respectively; 
thus, they are clearly accretors.

The star Kiso\,A-0974~17 was first classified as a H$\alpha$
emission-line object by \citet{Nakano95}. Although \citet{Alcala04}
did not obtain any spectrum of this source, they included it in
their list of PMS stars. We observed this object during the 
FLAMES spectroscopic run. Neither H$\alpha$ emission nor
Li\,{\sc i} absorption line were observed in its spectrum.
This could be due to a mismatch between the H$\alpha$
emission source and the optical counterpart detected by
\citet{Nakano95}. Thus we excluded this object from our list
of PMS stars.

The optical ($UBVR_CI_C$) and near-infrared ($JHKs$) magnitudes
of the 56 optically confirmed PMS stars in L1615/L1616 is
presented in Table~\ref{Tab:phot}. Some of the objects
lack WFI, OAC, and/or SOFI photometry because they fall outside 
the sky-area covered by our survey, are saturated in our images 
or are too faint to be observed with the 91\,cm telescope at 
the OAC observatory. Nevertheless, for these stars we 
retrieved $I_C$ magnitudes from the TASS Mark IV photometric 
survey \citep{Droege06}, $I_CJ$ magnitudes from the DEep 
Near-Infrared Survey \citep[DENIS;][]{Epchtein97}, and $JHKs$ 
photometry from the 2MASS point source catalogue. Both the SOFI 
and DENIS NIR magnitudes were homogenised to the 2MASS photometric 
system using the transformation equations provided by 
\citet{Carpenter01}. For a few sources, we used the optical photometry 
from the literature \citep{Lee68,Mundt80,Taylor89,Alcala96,Cieslinski97,
Frasca03,Vieira03} as explained in the footnotes of Table~\ref{Tab:phot}.

\clearpage
\clearpage

\section{Data analysis}
\label{sec:Data analysis}

\subsection{H$\alpha$ and Li\,{\sc i} equivalent widths}
\label{sec:Equivalent-Widths}

H$\alpha $ and Li\,{\sc i} $\lambda$6708~{\AA} line equivalent
widths, W(H$\alpha$) and W(Li), respectively, were determined
by direct integration of the line profiles, as described in
\citet{Alcala04} and \citet{Spezzi08}, adopting the convention that
a negative equivalent width means an emission line. The main
source of error on these measurements comes from the uncertainty
in the placement of the photospheric continuum, especially for
stars later than M1, whose spectra are strongly affected by
molecular absorption bands. W(H$\alpha$) and W(Li) for the 
L1615/L1616 members are reported in the third and fourth 
columns of Table~\ref{Tab:SpT-W}, respectively.

According to the intensity of the H$\alpha$ emission line and the
definitions by \citet{White03}, we assigned the classical
(CTTS) or weak (WTTS) T Tauri star classification to the PMS
objects in L1615/L1616 (Table~\ref{Tab:SpT-W}).

The additional emission lines observed in the spectra, including 
the lines of He\,{\sc i} $\lambda$5876, $\lambda$6678.7 and 
$\lambda$7065.2~{\AA}, and Na\,{\sc i} $\lambda$5889.9 and 
$\lambda$5895.9~{\AA}, as well as the forbidden lines of 
[N\,{\sc ii}] $\lambda$6583.5~{\AA}, [O\,{\sc i}] $\lambda$6300.3 
and $\lambda$6363.8~{\AA}, and [S\,{\sc ii}] $\lambda$6716.4 and 
$\lambda$6730.8~{\AA}, are also reported in Table~\ref{Tab:SpT-W}. 
Note that almost only those PMS stars classified as CTTSs show 
forbidden emission lines in their spectra.

\subsection{Spectral types and effective temperatures}
\label{sec:Spec-Type-Class}

Spectral types were determined using a modified version of
ROTFIT, a code for spectral type and $v\sin i$ determination
developed by \citet{Frasca03,Frasca06} under the
IDL\footnote{IDL is distributed by ITT Visual Information
Solutions, Boulder, Colorado.} environment. The general idea of
the method, already adopted in \citet{Alcala06} and
\citet{Spezzi08}, is to recover the spectral type of a star by
comparing its spectrum with a suitable grid of templates,
taking into account the amount of extinction along the line of
sight. Since our spectra are flux calibrated, the method allowed
us to get a rough estimate of the extinction ($A_{\mathrm V}$)
along the line of sight to the star.

To this aim, a library of relative flux calibrated and
extinction-corrected templates, with a wavelength coverage
encompassing that of the observed spectra, is needed. A suitable
grid of intermediate-resolution ($R\approx1000-5000$) templates
has been obtained by collecting spectra of dwarf and giant
stars from the libraries provided by
\citet{Martin99}, %($R\sim700$),
\citet{Hawley02}, %($R\sim1500$),
\citet{LeBorgne03}, %($R\sim2000$),
\citet{Valdes04}, %($R\sim3500$),
and \citet{Bochanski07}. %($R\sim1800$).
%All these templates are intermediate resolution spectra in the
%range $R\approx1000-3000$.

In performing the spectral type classification, we first took
into account the PMS nature of the stars we were dealing with.
Due to the presence of TiO and VO absorption bands which are
strongly sensitive to the surface gravity \citep{Torres-Dodgen93}, 
most of M-type PMS objects show optical spectroscopic features 
that can be closely reproduced averaging dwarf and giant spectra 
with the same spectral type \citep{Luhman99,Luhman03,Guieu06}. 
Though the spectra of G-type and K-type PMS stars have been usually 
compared with template of normal dwarf stars in several works 
\citep[e.g.][]{Basri90}, we found that also the 
late K-type objects in our sample are best represented by using an 
average of dwarf and giant templates of the same spectral type.
Indeed TiO and VO absorption molecular bands are still present 
in late K-type stars, although they are very weak.

Thus, both late-type dwarf and giant templates were included in our
grid of reference spectra; moreover, for spectral types later or equal 
than K5, we also built up an \emph{ad hoc} grid of templates by averaging
dwarf and giant spectra of the same spectral sub-class. Furthermore,
starting from spectral types later than K7, we constructed templates
for the half sub-classes (e.g. K8.5, M0.5, M1.5, etc.) by averaging
the two contiguous spectra. Indeed, the intensity of the absorption
molecular bands characterising late-type stars displays a rather
strong variation from one spectral sub-class to the other, allowing
to appreciate differences of half a sub-class. Some details on
the spectra fitting procedure is provided in
Appendix~\ref{App-A}.

The spectral types of the PMS objects in our sample are reported in 
the second column of Table~\ref{Tab:SpT-W}. From an inspection of the
residuals of the fitting procedure we have estimated an accuracy
of about $\pm\,1$ sub-class for stars earlier than K7 and about
$\pm\,0.5$ sub-class for cooler objects. A typical uncertainty on
$A_{\mathrm V}$ of $\pm\,0.2$~mag was found. For all the objects
observed at least twice, with different spectrographs and/or resolving
powers, we obtained consistent results, regardless of spectral
resolution and wavelength coverage. 
For the sake of homogeneity, we re-determined the spectral types also
for the previously known low-mass members, by using the spectra from
\citet{Alcala04}. A general agreement within one/two spectral 
sub-classes was found between our classification and the one 
reported in the literature \citep{Cohen79,Alcala04}.

For the two early-type stars HD\,293815 and Kiso\,A-0974~15,
we have adopted the spectral types from \citet{Sharpless52}
and \citet{Vieira03}, respectively. Both stars were
spectroscopically observed by us, but many photospheric features
suitable for an accurate early-type classification were not covered
in our spectra. Nevertheless, we attempted a spectral type estimate
for both stars by applying the \citet{Hernandez04}'s criteria. The
main features we used are the He\,{\sc i} $\lambda$5876 and
$\lambda$7065.2~{\AA} absorption lines. The results are consistent
with those reported by \citet{Sharpless52} and \citet{Vieira03}.
% %, based on the measurement of the equivalent width of suitable
% %photospheric features which are sensitive to changes in effective
% %temperature ($T_\mathrm{eff}$).

In line with our spectral type classification, the effective temperature
($T_\mathrm{eff}$) of each PMS star was assigned on the basis of the
following criteria: the dwarf temperature scale
from the compilation of \citet{KenyonHartmann95} was adopted for all
the objects with spectral type earlier or equal to K4; 
for later spectral types an intermediate temperature
scale was derived by averaging the giant compilation provided by
\citet{vanBelle99} with the dwarf one from \citet{KenyonHartmann95}
for K0-M0 and from \citet{Leggett96} for M1-M7 spectral types, respectively.
The $T_\mathrm{eff}$ values corresponding to the half spectral
sub-classes were computed by linear interpolation.
The effective temperatures assigned to each PMS object are reported
in the second column of Table~\ref{Tab:Param}. In assessing the errors
on $T_\mathrm{eff}$ the uncertainty on the spectral type classification
was taken into account.

\subsection{Extinction, stellar radii, and luminosities}
\label{sec:IntExt-Rad-Lum}

The method we used for computing %interstellar/circumstellar
extinction, stellar radii, and luminosities follows the general
guidelines described in \citet{Romaniello02,Romaniello06},
although we modified their scheme to match our purposes.
The eight broad-band magnitudes available for the L1615/L1616
PMS population allowed us to construct the spectral energy
distribution (SED) of each PMS star covering a wide spectral
range, from optical to near-infrared wavelengths
(see Table~\ref{Tab:abs-flux-calib}). By using simultaneously
all the photospheric colours encompassed by the SED, we
computed for each object both the interstellar extinction 
($A_{\mathrm V}$) and the ratio of total-to-selective extinction
($R_{\mathrm V}=A_{\mathrm V}/E_{\mathrm {B-V}}$), as well as
the stellar radius ($R_{\star}$) and luminosity ($L_{\star}$).

Apart for eventual ultraviolet and infrared excesses caused by
the presence of an interacting accretion disk around the PMS
star and by the related boundary layer, or hot accretion spots,
the shape of the observed SED is mostly determined by the
effective temperature of the object, the amount of dust along
the line of sight, and the dust-grain mean size in the
interstellar/cicumstellar environment.
Therefore, once the effective temperature is known, both the
value of $A_{\mathrm V}$ and $R_{\mathrm V}$ may be recovered
simultaneously by fitting the observed SED with a grid of
theoretical ones. Indeed, both the star radius and distance
cause just a rigid shift of the flux on a logarithmic scale,
without affecting the shape of the SED. Theoretical SEDs may
be obtained by using stellar atmosphere models with the same
effective temperature as the star and reddened by various
amount of $A_{\mathrm V}$ and $R_{\mathrm V}$.

A detailed description of the SED fitting procedure is
reported in Appendix~\ref{App-B}. Here we want to stress
that, our two-parameter fitting procedure can be applied
only to the objects which are significantly affected by
extinction, i.e to the stars which are still deeply embedded
in their parent cloud. Indeed, for slightly reddened stars the
value of $A_{\mathrm V}$ is relatively insensitive to changes
in the slope of the reddening law ($R_{\mathrm V}$), as shown
in Appendix~\ref{App-B}. Therefore, for all the objects for
which $A_{\mathrm V}$ turned out to be less than about 0.5 mag
we reiterated the fitting procedure fixing $R_{\mathrm V}$ to
the standard value of 3.1. The same assumption was adopted 
for all the objects lacking optical photometry and 
for the veiled star TTS\,050649.8$-$032104.

The values of $A_{\mathrm V}$, $R_{\mathrm V}$, $R_{\star}$,
and $L_{\star}$ are listed in Table~\ref{Tab:Param}.
For all the objects spectroscopically observed in this work,
we found that the value of $A_{\mathrm V}$ derived from the SED
fitting procedure is in agreement, within $\sim0.2$~mag, with
the one obtained from the spectral type classification.

$R_{\mathrm V}$ was derived for 15 out of the 56 PMS stars in
L1615/L1616 (Table~\ref{Tab:Param}). We found a weighted mean
value of $3.32\pm0.04$, closely matching the standard 
one of 3.1, typical of the diffuse interstellar medium.
For the Herbig star Kiso\,A-0974~15 we obtained a value of
$R_{\mathrm V}=5.5$ (see Table~\ref{Tab:Param}). This may
be related to the presence of dust grains around the
intermediate-mass star which are larger in size than those
characteristic of the interstellar medium, as already found
by different authors in large samples of HAeBe stars
\citep{The81, Herbst82, Waters98, Whittet01, Hernandez04}.

\subsection{Masses and Ages}
\label{sec:Masses-Ages}

\begin{figure}
  \centering
  \resizebox{\hsize}{!}{\includegraphics[draft=false]{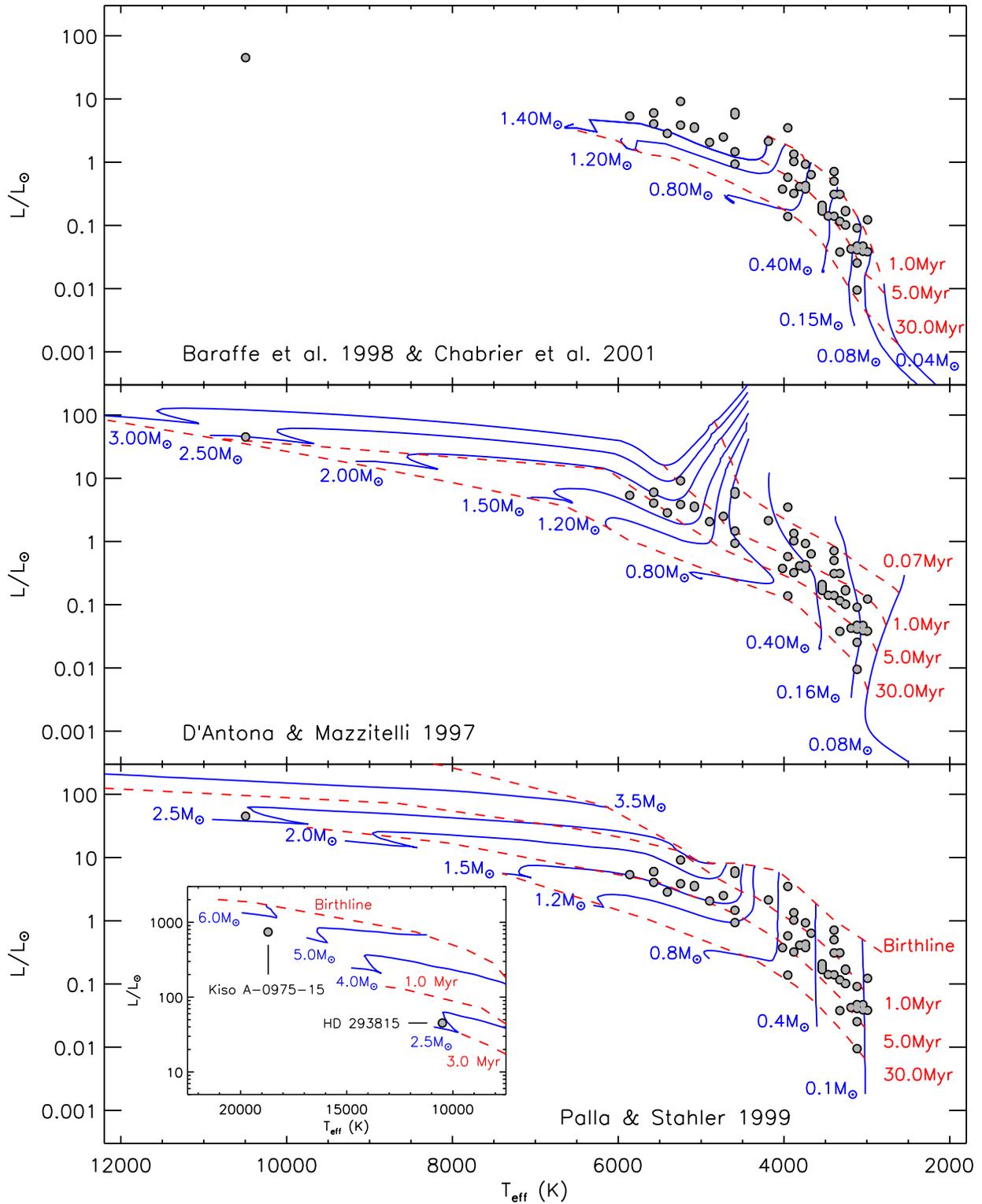}}
  \caption{H-R diagram for the young population in L1615/L1616; the PMS
           evolutionary tracks by \citet{Baraffe98} \& \citet{Chabrier00}
           (upper panel), \citet{Dantona97} (central panel), and \citet{Palla99}
           (lower panel) are over-plotted. The H-R diagram for the intermediate-mass
           members of L1615/L1616, namely HD\,293815 and Kiso\,A-0974~15, is shown
           in the inset of the lower panel.}
  \label{HR}
\end{figure}

The mass and age of each member of L1615/L1616 have been derived by
comparing the location of the object on the H-R diagram with the
theoretical PMS evolutionary tracks by \citet{Baraffe98} \&
\citet{Chabrier00}, \citet{Dantona97}, and \citet{Palla99}
(Figure~\ref{HR}).

Since stellar evolutionary models are still rather uncertain,
particularly in the low-mass star and sub-stellar regimes
\citep[$M<0.5 M_{\odot}$,][]{Baraffe02}, the use of different
evolutionary tracks allowed us to estimate the model-dependent 
uncertainties associated with the derived stellar parameters. 

Table~\ref{Tab:mass_age} lists masses and ages of the
L1615/L1616 members as inferred by the three sets of models.
Note that the evolutionary tracks by \citet{Baraffe98} \&
\citet{Chabrier00}, \citet{Dantona97}, and \citet{Palla99} are
available in the mass range $0.003\leq M \leq 1.40$~$M_{\odot}$,
$0.017\leq M \leq 3$~$M_{\odot}$, and $0.1\leq M \leq 6$~$M_{\odot}$,
respectively. Note also that isochrones by \citet{Baraffe98} 
\& \citet{Chabrier00} are available only for ages greater or 
equal than 1~Myr.

\clearpage
\clearpage

\section{Discussion}
\label{sec:Discussion}

Once the physical parameters of the PMS objects are known, we can
study the star formation in L1615/L1616. In the next sub-sections
we discuss the history, rate and efficiency of star formation,
and some issues related to the mass-spectrum.

\subsection{The star formation history}
\label{sec:Star-Form-Hist}

\clearpage
\begin{figure*}
  \centering
  \resizebox{\hsize}{!}{\includegraphics[draft=false]{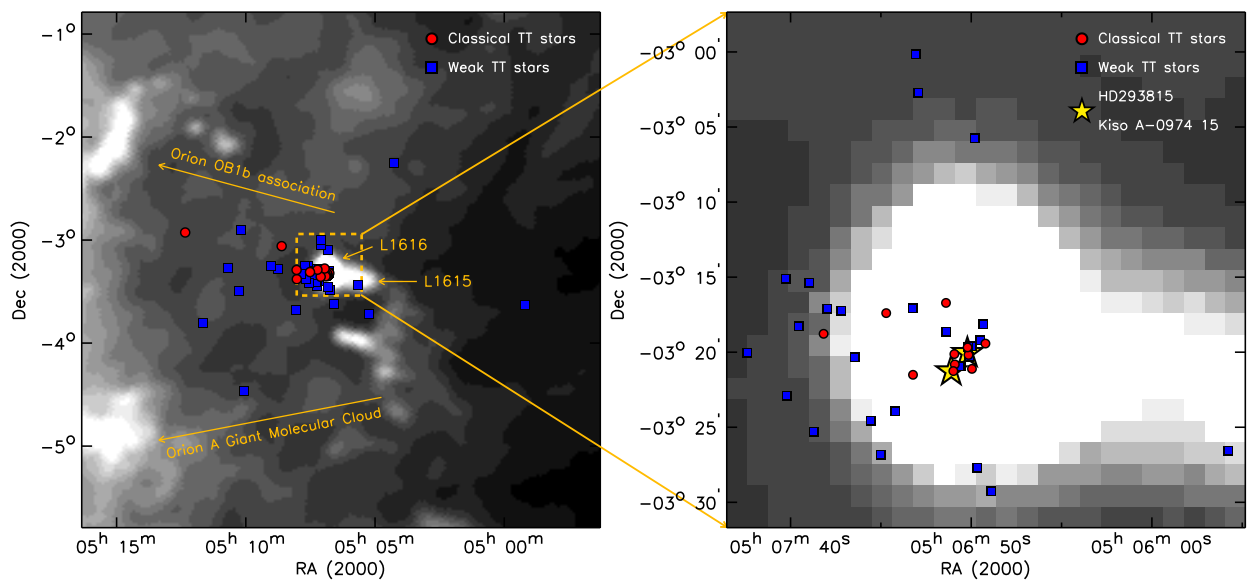}}%[height=6in,width=6in]
  \caption{IRAS 100$\mu$m dust emission map  covering a sky-area of $5^\circ\times5^\circ$ around L1615/L1616 
           (left panel). The Classical and Weak T Tauri stars are represented with filled dots and 
	   squares, respectively. The upper arrow shows the direction to the Orion OB1 associations, located 
	   at about 7.5$^\circ$ ($\sim60$ pc at a distance of 450 pc) to the North-East of L1615/L1616. The 
	   lower arrow points toward the Orion A Giant Molecular Cloud, also located at about 7.5$^\circ$ to 
	   the South-East of L1615/L1616. The dashed square defines the area surveyed with WFI. This region is 
	   zoomed in the right panel, where the two intermediate-mass members HD\,293815 and Kiso\,A-0974~15 
	   are marked with five-pointed star symbols.}
  \label{Fig:IRAS-Image}
\end{figure*}
\clearpage

Based on the most complete census of the L1615/L1616's population
performed in this work, we further investigated the triggered star
formation scenario suggested by \citet{Ramesh95}, \citet{Stanke02},
and \citet{Alcala04}.

In left panel of Figure~\ref{Fig:IRAS-Image} the IRAS~100~$\mu$m 
dust emission map of a $5^\circ\times5^\circ$ sky-area around 
L1615/L1616 is shown. About $64$\,\% of the CTTSs are clustered, 
in the densest part of L1615/L1616, i.e.~within the boundaries of 
the NGC\,1788 reflection nebula and to the West of the bright rim 
of the cloud (Figure~\ref{Fig:IRAS-Image}, right panel). Such rim 
is located to the East of NGC\,1788, at $\sim6.5{\arcmin}$ from the 
head of the cometary cloud (about 0.85 pc at a distance of 450 pc) 
and it is directly exposed to the UV radiation from the Orion OB 
stars. On the other hand, the WTTSs are more scattered (only $\sim22$\,\%
is projected on NGC\,1788) and mainly occupy the side of the cloud 
facing the OB1 associations and to the East of the bright rim.

By applying the Kolmogorov-Smirnov test to the WTTS and CTTS age 
distributions we found that the two populations are similar at a 
confidence level of $\sim 80$~\%. Thus, it seems there are no 
age differences between CTTSs and WTTSs in our PMS sample. 
This result is consistent with previous PMS populations studies 
\citep[see][and reference therein]{Feigelson99}; CTTSs are predicted to 
have ages between about 0.5 and 3 Myr, although some stars retain CTTS 
characteristics even at ages as old as 20~Myr. On the other hand, many 
WTTSs occupy the same region on the H-R diagram as CTTSs do, whereas 
some of them are approaching the zero-age main sequence.

\begin{figure}
  \centering
  \resizebox{\hsize}{!}{\includegraphics[draft=false]{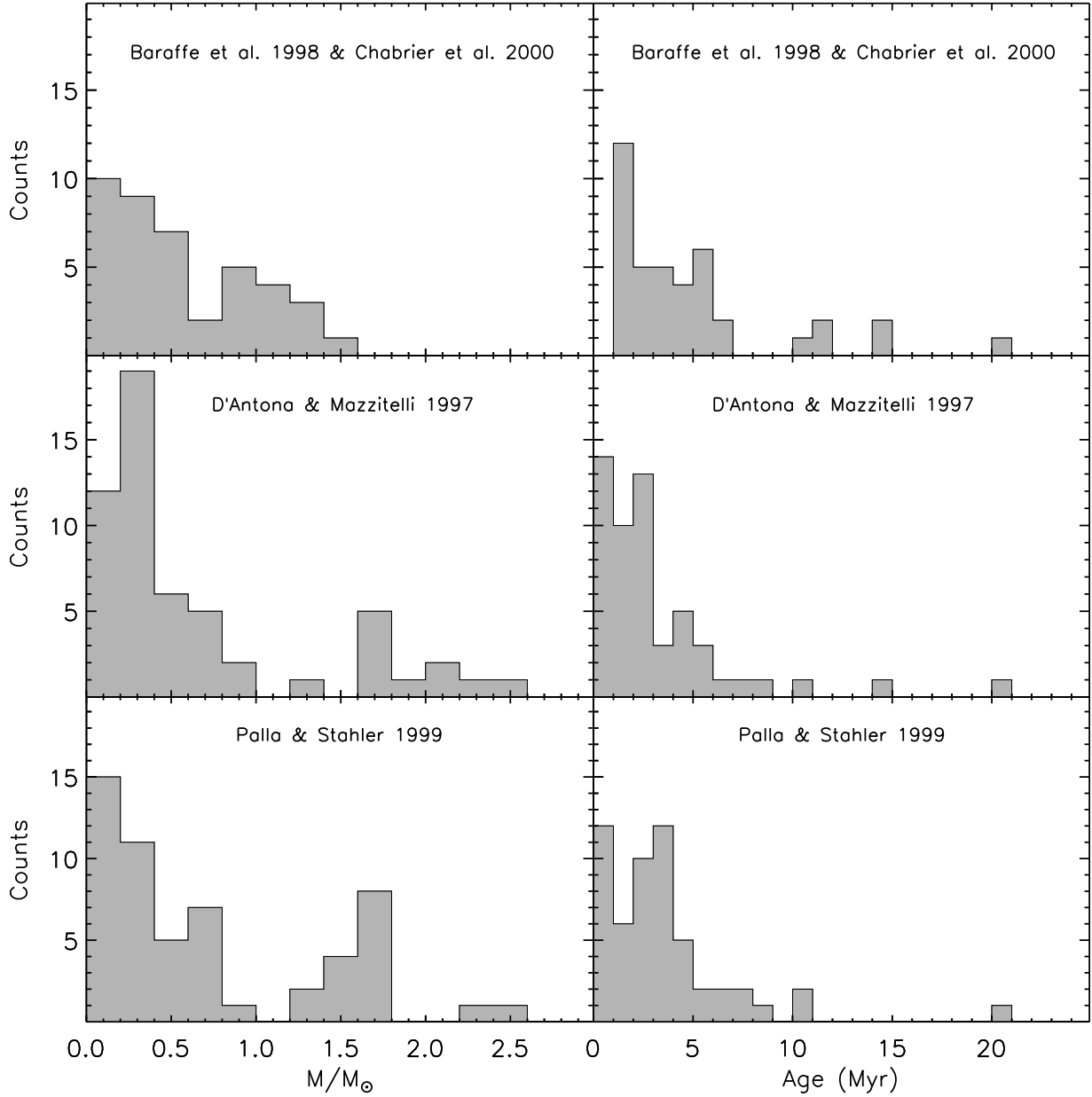}}
  \caption{Mass (left panel) and age (right panel) distributions of the L1615/L1616 PMS population derived by using the evolutionary tracks
           by \citet{Baraffe98} \& \citet{Chabrier00} (upper panels), \citet{Dantona97} (central panels) and \citet{Palla99} (lower panels).}
  \label{Fig:Hist_Ages_Masses}
\end{figure}

The age distribution of the L1615/L1616 population peaks between
1 and 3 Myr, depending on the adopted evolutionary tracks
(Figure~\ref{Fig:Hist_Ages_Masses}, right panels). This is in agreement
with the findings by \citet{Alcala04}, but the age spread found by us
is significantly higher than what these authors claim, exceeding the
value expected on the basis of the uncertainties on luminosity and
temperature. The L1615/L1616 members span a wide range in age, from less
than 0.1~Myr up to about 30~Myr. This might suggest multiple events of 
star formation in the cloud, which would further support the hypothesis 
of triggered star formation.

\begin{figure}
  \centering
  \resizebox{\hsize}{!}{\includegraphics[draft=false]{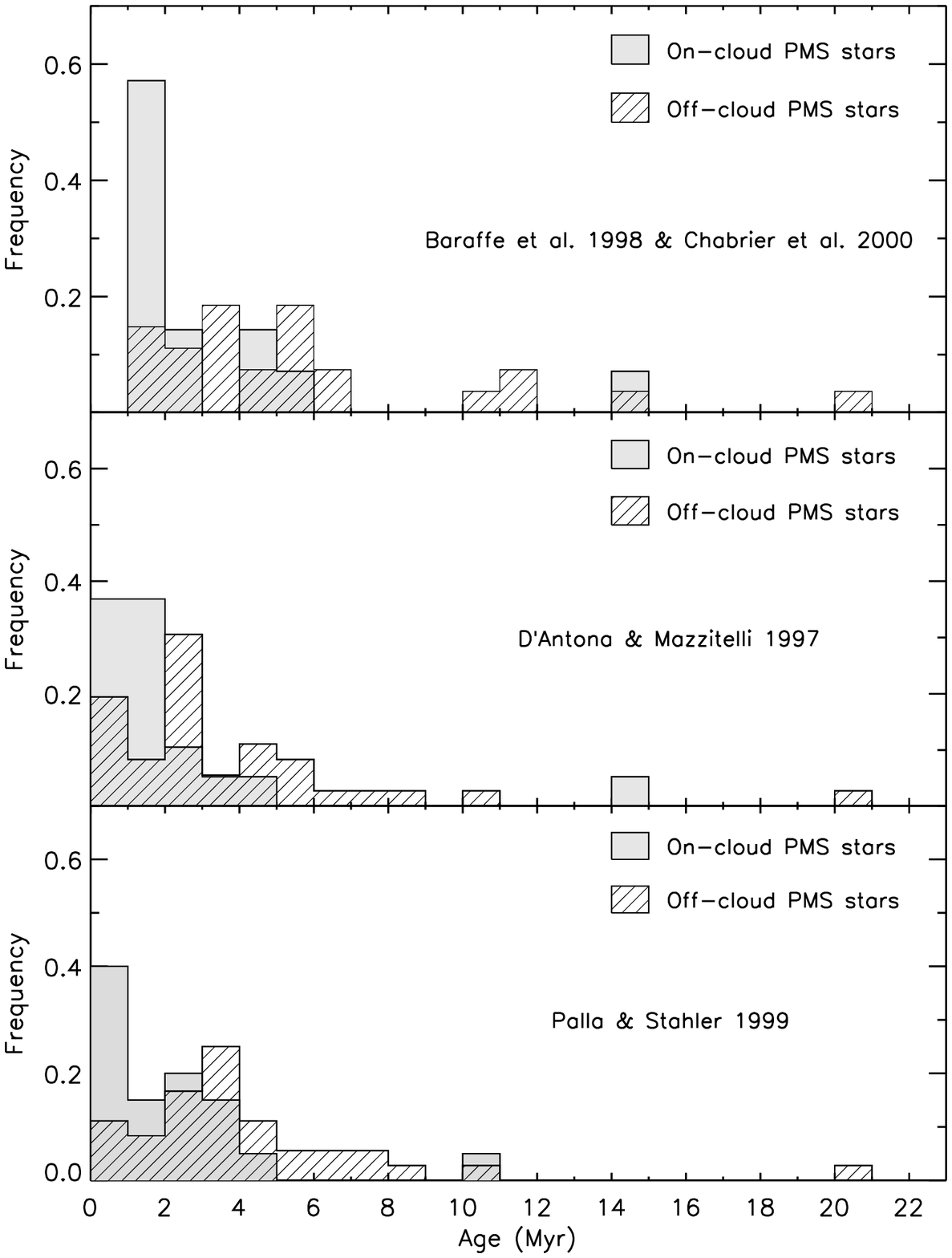}}
  \caption{Age distributions of the on-cloud and off-cloud PMS stars in L1615/L1616 derived by using the evolutionary
           tracks by \citet{Baraffe98} \& \citet{Chabrier00} (upper panel), 
	   \citet{Dantona97} (central panel) and \citet{Palla99} (lower panel).}
  \label{Compare_Age}
\end{figure}

In order to investigate a possible age difference between ``on-cloud''
and ``off-cloud'' PMS objects, we divided the sample in two groups,
fixing as dividing line the bright rim of the cloud.
Twenty of 56 objects in our sample are located within the  
boundaries of NGC\,1788, i.e. to the West of the bright rim, while 
36 are off-cloud members. The age distributions of the two groups 
are shown in Figure~\ref{Compare_Age}. The on-cloud PMS stars are 
statistically younger than those located to the East of the bright rim, 
regardless of the adopted evolutionary tracks. Applying a Kolmogorov-Smirnov 
test we found that the probability that the two sets are sub-sample 
of the same statistical population is very low; in particular, we 
found confidence levels of 1.29, 0.18, and 0.19\,\% when using the 
PMS evolutionary tracks by \citet{Baraffe98} \& \citet{Chabrier00}, 
\citet{Dantona97}, and \citet{Palla99}, respectively. We thus concluded 
that there is a clear age difference between the two groups of stars.

The above findings further support the scenario of triggered star
formation in L1615/L1616, as proposed by \citet{Stanke02}.
In this context, the spatial dispersion and older age of the ``off-cloud''
members can be explained as a consequence of the ``rocket acceleration''
effect. As recently pointed out by \citet{Kun04}, this acceleration continues
after the onset of star formation and the parental cloud is further accelerated
with respect to the newly formed objects. As a consequence, the cloud is soon
swept off the newly formed stars.
This hypothesis, together with the rapid dispersion typical of small clouds,
causes the spatial displacement of the oldest cloud members. The conspicuous
number of PMS stars found apparently isolated from classical star forming
regions \citep[e.g.][]{Alcala95,Covino97,Guillout98a,Guillout98b,Frasca03,
Zickgraf05} might be a consequence of this mechanism as well.

\clearpage

\clearpage
\thispagestyle{empty}
\setlength{\voffset}{-15mm}
\clearpage
\setlength{\voffset}{0mm}

\subsection{The star formation efficiency}
\label{sec:SFE}

Based on $^{12}$CO and $^{13}$CO column density maps, \citet{Ramesh95}
estimated that the mass of L1616 alone is in the range $169-193$~$M_{\sun}$,
depending on the used tracer. \citet{Yonekura99} mapped both L1615
and L1616 in the CO J=1-0 transition line. They inferred that the mass of
the cloud system as a whole is $\sim 530$~$M_{\sun}$, $\sim 350$~$M_{\sun}$
and $\sim 440 ~M_{\sun}$ based on $^{12}$CO, $^{13}$CO and $^{18}$CO
observations respectively. They also derived the mass of the $^{13}$CO
and $^{18}$CO cores (i.e. L1616 only) finding a value of $\sim
146$~$M_{\sun}$ and $\sim 161$~$M_{\sun}$, in good agreement with the
\citet{Ramesh95}'s determination.

\citet{Alcala04} estimated the SFE in L1616 to be $\sim14$\,\%, i.e.
higher than the average value measured in other low-mass star forming region
($\la3$\,\%). The SFE reported by \citet{Alcala04} for L1616 is based
on the cloud mass reported by \citet{Ramesh95} and the total stellar
mass of the 32 members of the cloud investigated by the authors
(i.e. $\sim30~M_{\odot}$).

Since the star formation history of L1615 and L1616 is intimately
connected, we have re-calculated the SFE considering the system L1615
plus L1616 as a whole. By using the NICER color excess method by
\citet{Lombardi01}, we mapped the dust extinction of the cloud complex
and derived its mass. In the NICER technique the $J-H$ and $H-Ks$
colours of the field stars are compared to the colours of stars in a
nearby reference field. The colour-excesses of the field stars are 
then combined and transformed to the visual extinction $A_{\mathrm V}$, 
fixing the form of the reddening curve. The normal interstellar extinction 
law derived by \citep{Cardelli89} has been adopted by us for this 
purpose. We retrieved the 2MASS colours of all the point-like sources 
located in a $50{\arcmin}\times50{\arcmin}$ region centered on NGC\,1788 
and in a reference field close to the L1615/L1616 cloud\footnote{The SOFI 
survey is only centered on the densest part of the cloud and can not be 
used to map the cloud as a whole.}. The latter field was selected by 
using the nearby low-intensity regions of the IRAS~100$\mu$m dust 
emission map, according to prescription by \citet{Kainulainen06}. 
Finally, the visual extinction map resulting from the NICER method 
was converted to cloud mass by assuming the standard gas-to-dust 
ratio \citep{Bohlin78}. We found a value of $\sim550$~$M_{\odot}$, 
in good agreement with the one inferred by \citet{Yonekura99} on the 
basis of $^{12}$CO observations. Based on this value and on the total 
mass of the 56 present-day known members of the complex (i.e. 
41--46~$M_{\odot}$, depending on the adopted evolutionary tracks), we 
derived a SFE in L1615/L1616 of 7--8\,\%, i.e. significantly lower 
than the previous estimate by \citet{Alcala04}, but still in good 
agreement with the one generally found in giant molecular clouds 
hosting OB associations (5--10\,\%) and predicted by theoretical 
calculations on the formation of OB associations 
\citep[see][and reference therein]{Clark05}.

\subsection{Density of star formation and star formation rate}
\label{sec:SFR}

An interesting question is whether L1615/L1615 can be considered as
a cluster. According to the definition suggested by \citet{Lada03}, 
a cluster is a group of some 35 members with a total mass density 
larger than 1.0~M$_{\odot}$\,pc$^{-3}$.
To estimate the density of PMS objects in L1615/L1615, we considered the
region spectroscopically surveyed by us (Figure~\ref{Fig:SpecSurvey}).
This region covers an area of approximately 0.25 square degrees and includes
about 40 PMS objects. Assuming a distance of 450~pc, the resulting area is
about 4~pc$^2$, which means a surface density of about 10-11 PMS objects
per pc$^2$ and a volume density on the order of 7 PMS objects per pc$^3$.
In the latter calculation we estimated the volume as $V = 0.752{\times}Area^{1.5}$ 
\citep[see][]{Jorgensen07}, assuming a locally spherical distribution of sources.
The average mass of the 40 PMS objects, as determined from the results in
Section~\ref{sec:Masses-Ages}, is on the order of 0.7~M$_{\odot}$, 
which implies a volume density of about 4-5~M$_{\odot}$\,pc$^{-3}$.
Therefore, the group of 40 PMS objects confined in the spectroscopically
surveyed area can be considered as a cluster according to the criterion
by \citet{Lada03}.

Now, it is interesting to estimate the rate at which the stars in this
small cluster are formed. According to the results of Section~\ref{sec:Masses-Ages},
we estimated that the total mass in PMS objects in that region is on the
order of 41--46~M$_\odot$. Therefore, considering the average age
of 2~Myr for these objects (see Figure~\ref{Fig:Hist_Ages_Masses}), we found
that the cometary cloud is turning some 20-23~$M_{\odot}$ into PMS objects
every Myr, which is lower than the star formation rate in other clusters
like those in Serpens \citep{Harvey07}, but higher than in
other T~associations like Chamaeleon~II \citep{Alcala07} and Lupus \citep{Merin07}.
Thus we concluded that L1615/L1616 is a small cluster with a moderate star
formation rate.

\subsection{The Initial Mass Function}

Though based on a low-number statistics, the first clue on the Initial
Mass Function in L1615/L1616 was given by \citet{Alcala04}. Based on
this study, the IMF in this region appears roughly consistent with
that of the solar neighbourhood in the mass range $0.3<M<2.5~M_{\odot}$
\citep{Miller79}. The authors also found several candidates for young
BDs and estimated the expected number of BDs relative to more massive
PMS stars in L1615/L1616 to be intermediate between Taurus
\citep[$\sim13$\,\%,][]{Briceno02} and the Trapezium cluster
\citep[$\sim26$\,\%,][]{Luhman00}.

The depth of the spectroscopic survey conducted in this work 
goes down to $I_C\approx19.0$~mag (i.e. $M\approx0.03-0.05 ~M_{\odot}$, 
see Section~\ref{sec:BD-fraction}), leading to the discovery of young 
objects with a mass close to the Hydrogen burning limit. We have 
thus attempted a fair IMF determination down to the low-mass stellar 
regime (i.e. $M \gtrsim 0.1~M_{\odot}$).

All the present-day known members of L1615/L1616 have masses in the
range between $\sim$0.1 and 5.5~$M_{\odot}$; their mean mass varies
between 0.6 and 0.8~$M_{\odot}$ depending on the adopted evolutionary
tracks (see Table~\ref{Tab:mass_age} and Fig~\ref{Fig:Hist_Ages_Masses},
left panels). However, although our survey recovers most of the 
previously known young objects in L1615/L1616, the completeness of our 
PMS sample is different from the one of \citet{Alcala04} sample. The 
latter includes bright and solar-like objects located in areas not 
covered by the deeper WFI survey. Therefore, we restriced our IMF 
analysis to the sample of objects that lie on the sky-area covered 
with WFI (Figure~\ref{Fig:IRAS-Image}, right panel).

The shape of the IMF in the cloud may then be constrained using the usual 
approximation for the low-mass function ($\frac{dN}{dM} \propto M^{-\alpha}$), 
as reported by \citep{Moraux03}. We divided the mass range into mass bins 
of about $0.2~M_{\odot}$; this value is larger than the accuracy on mass
estimates as derived from the uncertainties on temperatures and luminosities
and, at the same time, allows us to have a statistically significant
number of objects in each bin.

\clearpage
\begin{figure}
 \resizebox{\hsize}{!}{\includegraphics[draft=false]{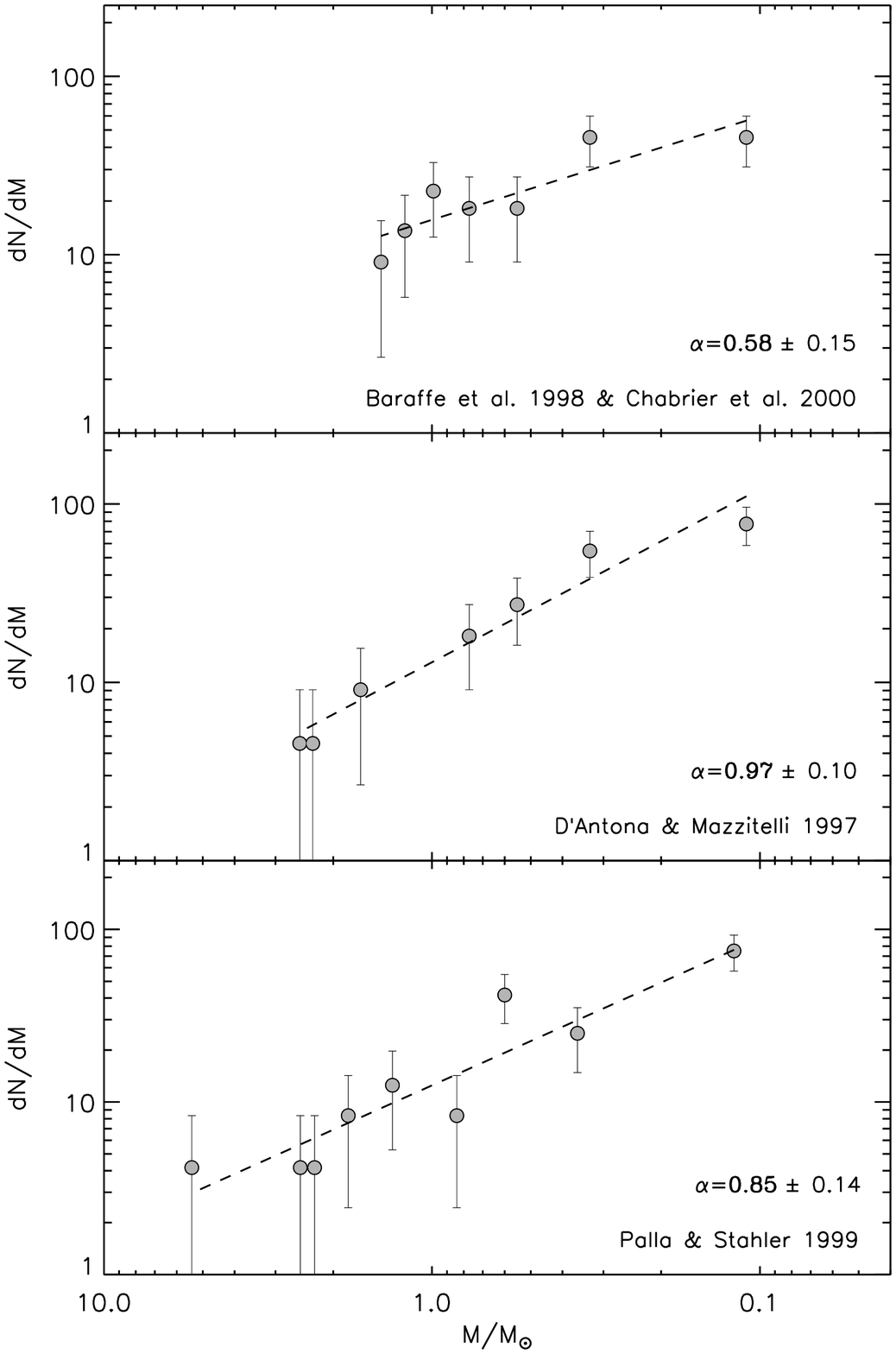}}
  \caption{The L1615/L1616 IMF between 0.1 and 5.5~$M_{\odot}$ as derived by
           using the evolutionary tracks by \citet{Baraffe98} \&
           \citet{Chabrier00} (upper panel), \citet{Dantona97} (central panel) and
           \citet{Palla99} (lower panel).}
  \label{Fig:IMF}
\end{figure}
\clearpage

Figure~\ref{Fig:IMF} shows the IMF in L1615/L1616 as obtained by using the
three sets of evolutionary tracks by \citet{Baraffe98} \& \citet{Chabrier00},
\citet{Dantona97} and \citet{Palla99}. Values of {\large $\alpha$} in the
three cases agree within the errors and we can derive a weighted mean
value of:
$$
 \alpha=0.84 \pm 0.07
$$
as the best approximation of the IMF slope in L1615/L1616 in the
mass range $0.1{\le}M{\le}5.5$~$M_{\odot}$. We notice, however, that the
evolutionary models by \citet{Dantona97} tend to produce a steeper IMF with
respect to those by \citet{Baraffe98} \& \citet{Chabrier00} and
\citet{Palla99}, as already found by \citet{Spezzi08}.

In Table~\ref{Tab:IMF} the $\alpha$ slope obtained for the IMF in
L1615/L1616 is compared with those obtained in other star forming regions
with different ages and environmental conditions. These values indicate
a common shape of IMF in the mass range from $\sim0.1$ to $\sim1~M_{\odot}$.
This provides further evidence of a universal IMF also
in the low-mass regime and is in agreement with the results already
established for the intermediate and massive star domains \citep{Kroupa02}.

\subsection{No sub-stellar objects in L1615/L1616\,?}
\label{sec:BD-fraction}

Prior to our survey, no confirmed BDs in L1615/L1616 were reported in
the literature. However, \citet{Alcala04} photometrically selected
some 30--40 candidates for BDs in this region and estimated a fraction
of sub-stellar objects relative to PMS stars ($R_{SS}$) in the range
from 18\,\% to 25\,\%, i.e. similar to the value measured in the
Trapezium cluster \citep{Hillenbrand00,Briceno02,Muench02} and in other
OB associations.

Our spectroscopic observations, which investigated about 50\,\% 
of the BD candidates with $I_C\lesssim19.0$ mag reported by 
\citet{Alcala04}, revealed 3--4 young objects with mass $M\le0.1~M_{\odot}$ 
(see Table~\ref{Tab:mass_age}), close to the Hydrogen burning limit, 
and no objects with lower mass. However, taking into account the 
accuracy with which we estimated the masses, these objects could be BDs 
with a mass just below the Hydrogen burning limit. The $R_{SS}$ in the 
cloud would be then around 5--7\,\%. Though the very low-mass BDs 
($M\la0.03~M_{\odot}$) and some deeply embedded low-mass objects might 
have escaped detection in our survey, this value should be slightly 
lower than the one estimated for other T~associations 
\citep[12--15\,\%;][]{Briceno02,Lopez04}.

The photometric survey conducted by us in L1615/L1616 is spatially and
photometrically complete down to $I_C\approx21.5$~mag at $10\,\sigma$
level. However, only objects brighter than $I_C\approx19.0$ mag could
be observed spectroscopically with a S/N ratio sufficient for spectral
type classification purpose. A 1--3~Myr old object (i.e. the typical 
age of the L1615/L1616 population) at the distance of 450~pc (i.e. a 
distance modulus of 8.27~mag) with a mass $M\ga0.03~M_{\odot}$ would 
have $I_C\la19.0$~mag in the absence of extinction \citep{Baraffe98}. 
Thus, in the off-cloud regions of L1615/L1616, where $A_{\mathrm V}\approx0$~mag, 
our survey is sensitive down to $\sim0.03~M_{\odot}$.
The visual extinction occurring in the dense head of the cloud varies in
the range 1--5~mag which displaces the limiting mass of the objects with
$I_C\la19.0$~mag to 0.05--0.50~$M_{\odot}$ \citep{Baraffe98}. We conclude 
that we are probably missing the very low-mass objects only in the
small dense core of the cloud. Given the peculiar star formation history
in this cloud (Section~\ref{sec:Star-Form-Hist}), the knowledge of its
sub-stellar mass spectrum could give crucial insights into the process
forming BDs under particular environmental conditions.

\clearpage
\begin{figure}
 \centering
 \resizebox{\hsize}{!}{\includegraphics[draft=false]{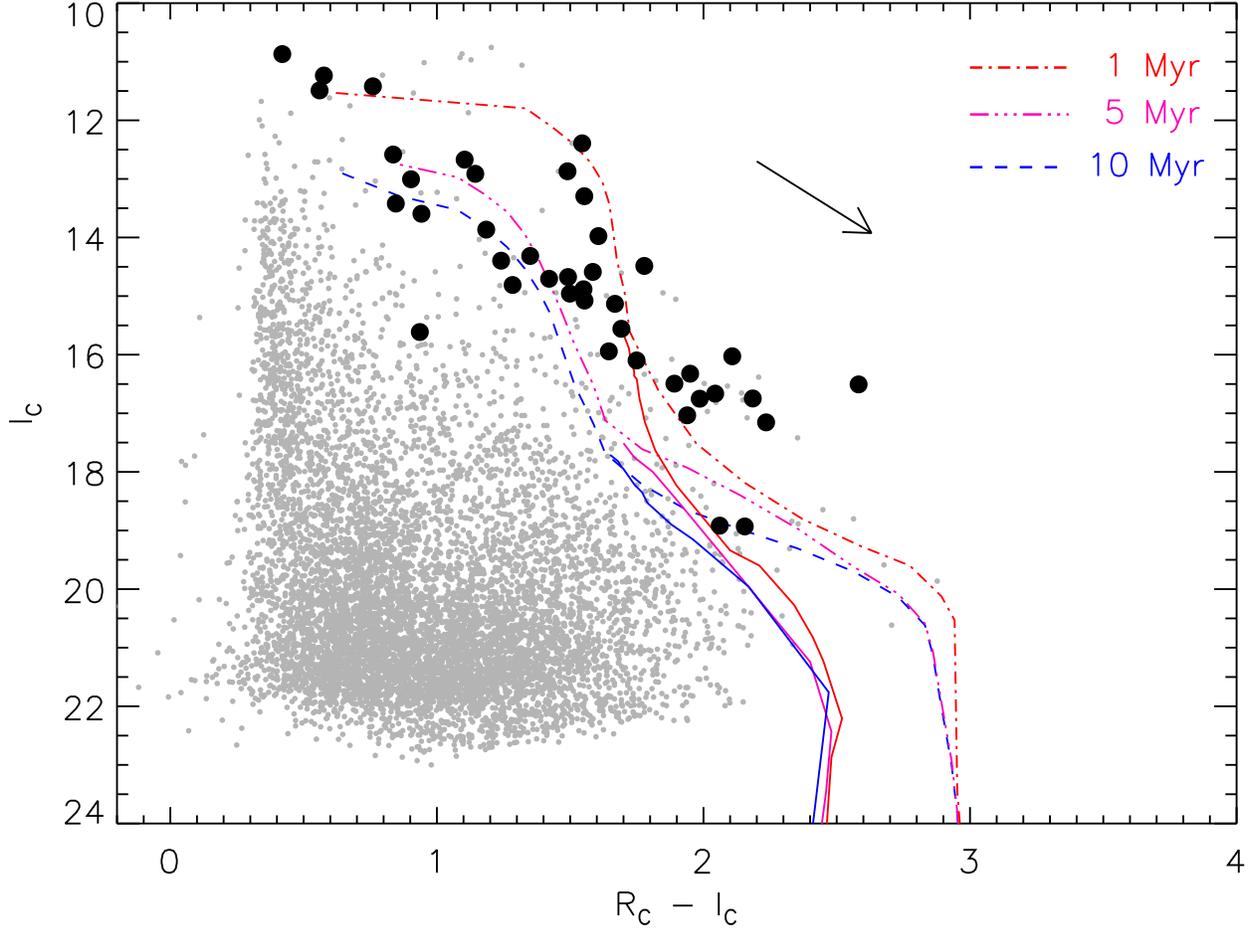}}
 \caption{$I_C$ versus $(R_C-I_C)$ diagram for the point-like objects in
          L1615/L1616 detected in our WFI survey above $3\,\sigma$ level 
          (grey little dots). The theoretical isochrones by \citet{Baraffe98} \&
          \citet{Chabrier00} for the Cousins photometric system of Bessel
          (continuous line) and for the WFI-Cousins system
          \citep[dashed lines,][]{Spezzi07} are overplotted. The PMS stars
          with both $R_C$- and $I_C$-band measurements (see Table~\ref{Tab:phot})
          are overplotted with black big dots. The $A_{\mathrm V}=2$~mag 
          reddening vector is also shown. The confirmed PMS object falling well 
          below the 10~Myr isochrone is the veiled star TTS\,050649.8$-$032104}.
 \label{Fig:isoc_Bes}
\end{figure}
\clearpage

The BD candidates sample by \citet{Alcala04} was selected by inspection
of $I_C$ versus $(R_C-I_C)$ colour-magnitude diagram in comparison with
the theoretical isochrones by \citet{Baraffe98} \& \citet{Chabrier00}.
This approach to search for PMS star and young BD candidates is the most
direct way to tackle the problem using optical photometric data; however,
in order to select cluster members on the basis of their magnitudes and
colours, isochrones that adequately model the data must be used, since
the shape of the isochrones may vary significantly from one photometric
system to another.
The theoretical isochrones for low-mass stars and BDs are provided by
\citet{Baraffe98} \& \citet{Chabrier00} in the Cousins photometric
system \citep{Cousins76}, while the observations by \citet{Alcala04}
were performed by using the WFI camera at the ESO 2.2m telescope.
The WFI filters transmission curves are somewhat different from the
original Cousins ones, in particular for the $I$-band. \citet{Spezzi07}
have computed \emph{ad hoc} isochrones for the WFI-Cousins system. In
Figure~\ref{Fig:isoc_Bes} these isochrones are compared with those by
\citet{Baraffe98} \& \citet{Chabrier00}. The two sets of isochrones
appear significantly different, in particular for the coolest objects,
i.e. $(R_C-I_C)>1.7$, where the use of the isochrones in the Cousins
system would produced many spurious member candidates. This
consideration, together with the contamination by background/foreground
stars in the field taken into account by the authors, could explain
why \citet{Alcala04} overestimated the number of BD candidates. The use
of the adequate isochrones for the WFI-Cousins system would produce
much less BD candidates, in agreement with the number of PMS objects
with mass close to the Hydrogen burning limit confirmed by our
spectroscopic follow-up (see Table~\ref{Tab:mass_age}).

Though the $R_{SS}$ estimated by us for L1615/L1616 needs further
refinements, the region seems to be poor of BDs, with a $R_{SS}$ ratio
close to that observed in T~associations or, perhaps, even lower.
Though in L1615/L1616 the radiation from the massive stars in the
Orion OB1 associations is likely responsible for the star formation
(Section~\ref{sec:Star-Form-Hist}), this primordial process does not seems
to favour the formation of BDs. As reported in \citet{Alcala04}, the
radial velocity (RV) distribution of the stars in L1615/L1616 shows
a well defined peak at $22.3$~Km/s with the standard deviation of
$4.6$~Km/s which is consistent with the average RV error (i.e. $\sim5$~Km/s).
This may indicate that the velocity dispersion of the stars must be
less than $5$~Km/s. Considering a velocity dispersion of a few Km/s,
$2$~Myr old objects would disperse over a distance of about $2$~pc,
which is the approximate projected size of the head of the cometary
cloud and has been completely covered by our photometric and
spectroscopic observations.

Thus, the results of our study in L1615/L1616 does not play in favour
of dynamical ejection or photoevaporation by ionising radiation from
massive stars being triggering factors of the BDs formation mechanism.

\section{Summary}

We have undertaken a study of the star formation process in the
L1615/L1616 cometary cloud, a small star forming region located in
the outskirt of the Orion OB associations. We aimed at characterising
its PMS population and studying both its IMF and star formation
history, in order to assess the role of the triggered star formation
scenario.

Our study is based on optical and near-infrared photometric observations,
as well as multi-object follow-up spectroscopy, both carried out
using ESO and OAC facilities. Complementary $JHKs$ photometric
data from the 2MASS and DENIS surveys has been used as well.

One of the major goal of this work has been the physical
parametrisation of the young stellar population in L1615/L1616. The
spectral type classification has been performed by using a grid of
reference spectra of giants, dwarfs, and intermediate templates, 
constructed by averaging spectra of giant and dwarf stars
of the same spectral type. We have computed stellar luminosities by
means of a SED fitting procedure properly developed by us, which also
evaluates the interstellar/circumstellar reddening. This allowed us
to derive the mass and age of each member of L1615/L1616 by comparing
the location of the object on the H-R diagram with different sets of
theoretical PMS evolutionary tracks.

Our analysis of the young population in L1615/L1616 yielded the 
following results:

\begin{itemize}
\item By using the H$\alpha$ emission line intensity, as well as the
      strength of the lithium $\lambda$6708~{\AA} absorption line
      as main diagnostics of the PMS nature of the objects, we
      identified 25 new members of L1615/L1616, almost doubling the
      number of previously known young objects in this cloud.
\item The age distribution of the L1615/L1616 population peaks
      between 1 and 3 Myr; however, the members of the cloud span a
      wide range in age, from $\sim$0.1~Myr up to 30~Myr,
      suggesting multiple events of star formation.
\item The evidence of multiple star formation events, the spatial
      distribution of the classical and weak T~Tauri populations,
      as well as the cloud shape, support the hypothesis of the star
      formation in L1615/L1616 being triggered by the massive stars
      in the nearby Orion OB associations.
\item The slope of the IMF in L1615/L1616 in the mass range
      $0.1{\le}M{\le}5.5$~$M_{\odot}$ is $0.84\pm0.07$, i.e. similar to that
      obtained in other star forming regions by different authors.
      This provides further support of a universal IMF in the stellar
      domain, regardless of the environmental conditions.
\item The star formation efficiency is about 7--8\,\% as expected for
      molecular clouds in the vicinity of OB associations.
      According to the criterion by \citet{Lada03}, L1615/L1616
      is a small cluster with moderate star formation.
\item A very low-fraction of possible sub-stellar objects ($\sim$~5--7\,\%)
      is found in L1615/L1616; thus, the dynamical interaction with the
      close massive stars in the Orion OB associations and/or the
      photoevaporation induced by their ionising radiation do not
      appear as efficient triggering factors for BD formation.
\end{itemize}


%% Included in this acknowledgments section are examples of the
%% AASTeX hypertext markup commands. Use \url without the optional [HREF]
%% argument when you want to print the url directly in the text. Otherwise,
%% use either \url or \anchor, with the HREF as the first argument and the
%% text to be printed in the second.

\acknowledgments

We thank the anonymous referee for his/her careful reading, useful 
comments, and suggestions which helped to improve the manuscript.
This paper is based on observations carried out at the European Southern
Observatory, La Silla and Paranal (Chile), under observing programs
numbers 64.I-0355, 70.C-0629, 70.C-0536, 074.C-0111 and 076.C-0385.
The authors are grateful to Menadora Barcellona and Salvatore Spezzi
for their supports during the preparation of this paper. We thank 
J.~Hernandez for providing the calibration curves for spectral
classification, and for discussions concerning the spectral
classification of early-type stars. We are also grateful to Matula for
her assiduous and warm assistance during the writing of this manuscript.
This work was partially financed by the Istituto Nazionale di Astrofisica
(INAF) through PRIN-INAF-2005. D.G. acknowledges financial support from
PRIN-INAF-2005 (``Stellar clusters: a benchmark for star formation and
stellar evolution''). This publication makes use of data products from the
Two-Micron All-Sky Survey, which is a joint project of the University
of Massachusetts and the Infrared Processing and Analysis Center/California
Institute of Technology, funded by NASA and the National Science Foundation.
This research has made use of the SIMBAD database, operated at CDS,
Strasbourg, France. This paper includes data from the DENIS project,
which has been partly funded by the SCIENCE and the HCM plans of the
European Commission under grants CT920791 and CT940627.


%% in parentheses, after the keyword, but they will not be verified.

% {\it Facilities:} \facility{WFI@2.2m}, \facility{SOFI@NTT}, \facility{FORS2@VLT, VIMOS@VLT, FLAMES@VLT}.



\appendix

\section{Spectral Type Classification}
\label{App-A}

In order to find the template spectrum which better reproduces the target
spectrum, we proceed as follows:

\begin{itemize}

\item We have first performed a spline interpolation of each
     standard spectra onto the wavelength points of the target
     spectrum. This allowed us to get a homogeneous wavelength
     grid with the same step as the observed spectrum. 
     Moreover, the spectral regions encompassing
     both the O$_{3}$ and H$_2$O telluric bands near 6900, 7250,
     and 7600~{\AA}, have been excluded in the fitting procedure.

\item The standard spectra have been shifted in wavelength to
      the radial velocity of the target; a cross-correlation algorithm
      was used to find the velocity shift between the templates and
      target spectrum and superimpose each other.

\item The extinction law derived by \citet{Cardelli89} have been adopted
      to redden the wavelength-shifted standard spectra before comparing
      them to the observed one. We assumed a normal slope
      of the reddening law, i.e. we adopted a value of the ratio of
      total-to-selective extinction $R_{\mathrm V}=A_{\mathrm V}/E(B-V)$
      equal to 3.1.
      Each standard spectrum has been progressively reddened assuming
      an increasing value of the extinction in the $V$-band
      ($A_{\mathrm V}$); we let it vary from 0 to 15 mag, with a step
      of 0.05 mag.

\item After each reddening step, the standard spectrum was subtracted
      to the observed one. The sum of the residuals was taken as
      indicator for the goodness of the ``fit''. The reddened standard
      spectrum which gave the lowest value of this indicator provided the
      spectral type of the target as well as a first estimation of the
      extinction value $A_{\mathrm V}$.

\end{itemize}

\begin{figure}
 \centering
 \includegraphics[width=10cm]{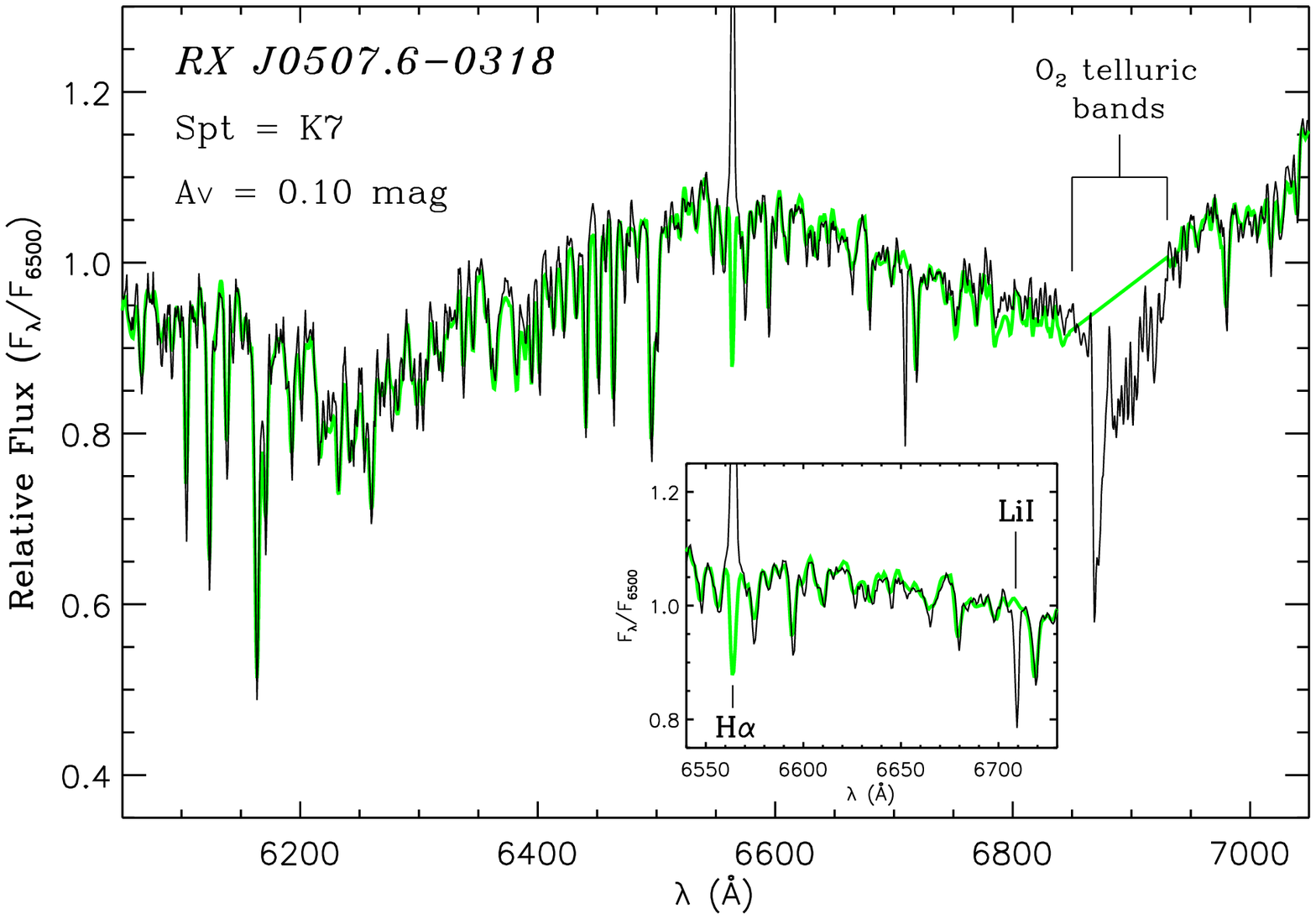}
 \includegraphics[width=10cm]{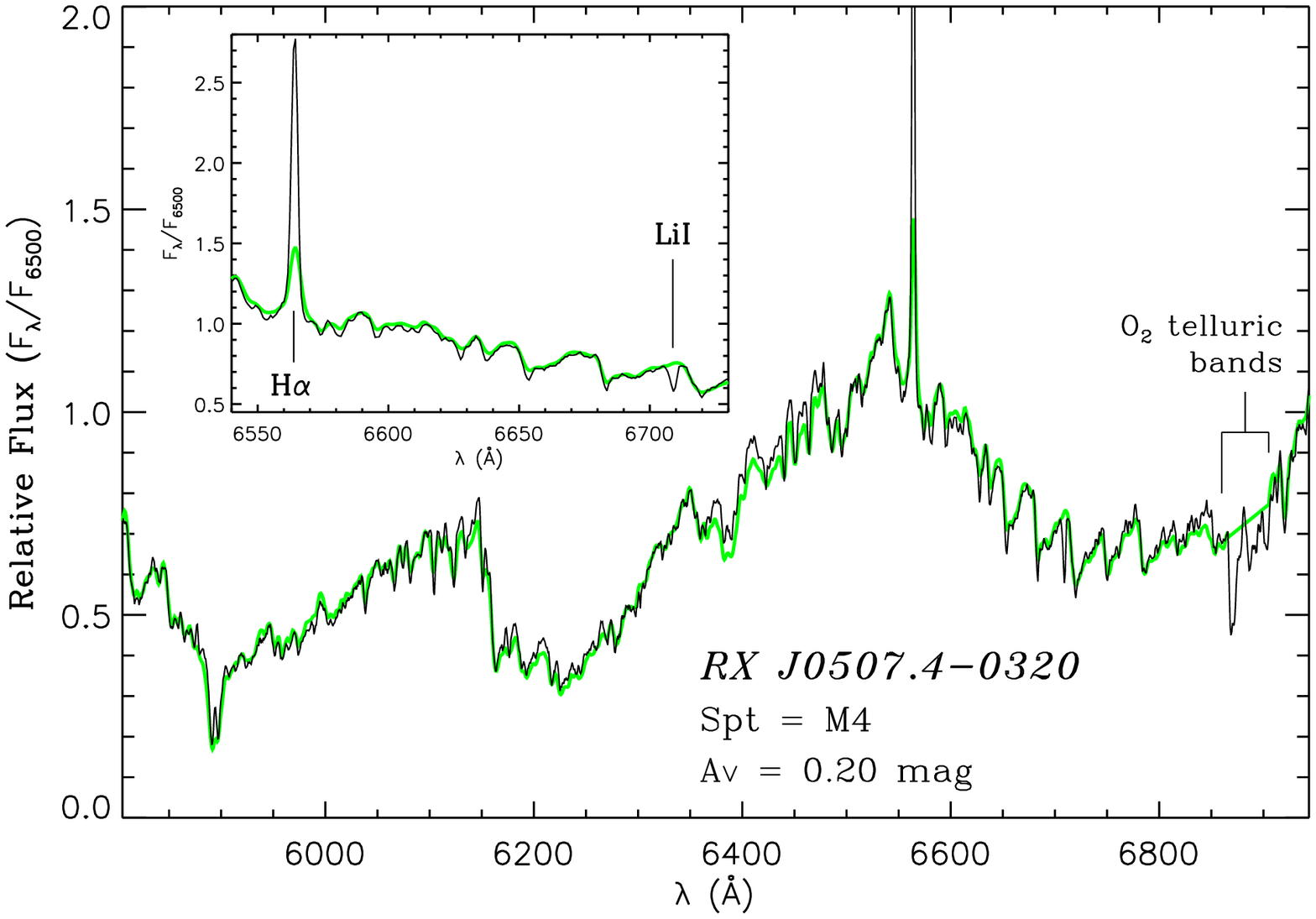}
 \caption{FORS2 spectra (thin lines) of RX\,J0507.6$-$0318
          (upper panel) and RX\,J0507.4$-$0320 (lower panel).
          Overplotted with thick lines are the best-fitting
          K7 and M4 templates, reddened with 
          $A_{\mathrm V}=0.1$ and $0.2$ mag, respectively,
          as derived by the spectrum fitting procedure.
          The spectra are arbitrarily normalised to the flux at
          6500~{\AA}. The O$_2$ absorption telluric bands near
          $\sim$6900~{\AA} are marked. This spectral region has
          been excluded in the fitting procedure. The Li\,{\sc i}
          $\lambda$6708~{\AA} absorption and the H$\alpha$ emission
          lines are shown in more detail in the inset of each panel. 
          Note the differences in H$\alpha$ and Li\,{\sc i}
          between the template and the observed spectra, confirming
          the PMS nature of the two stars. Also note the good
          agreement in the two independent determinations of
          $A_{\mathrm V}$, as obtained from the spectral type
          classification and the SED fitting procedure
          (Table~\ref{Tab:Param}).}
 \label{FORS2-spectra-fit}
\end{figure}

\begin{figure}
 \centering
 \includegraphics[width=10cm]{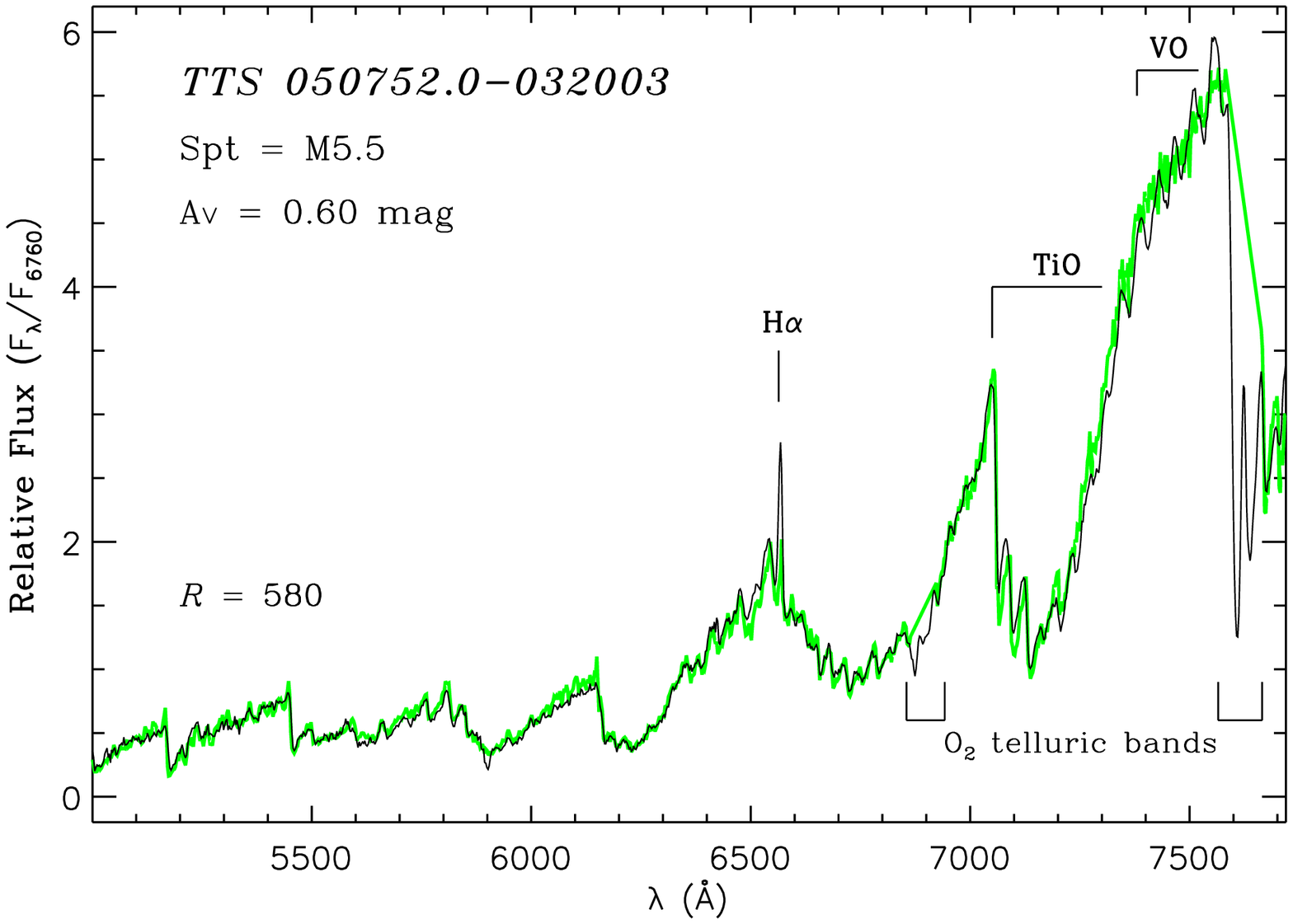}
 \includegraphics[width=10cm]{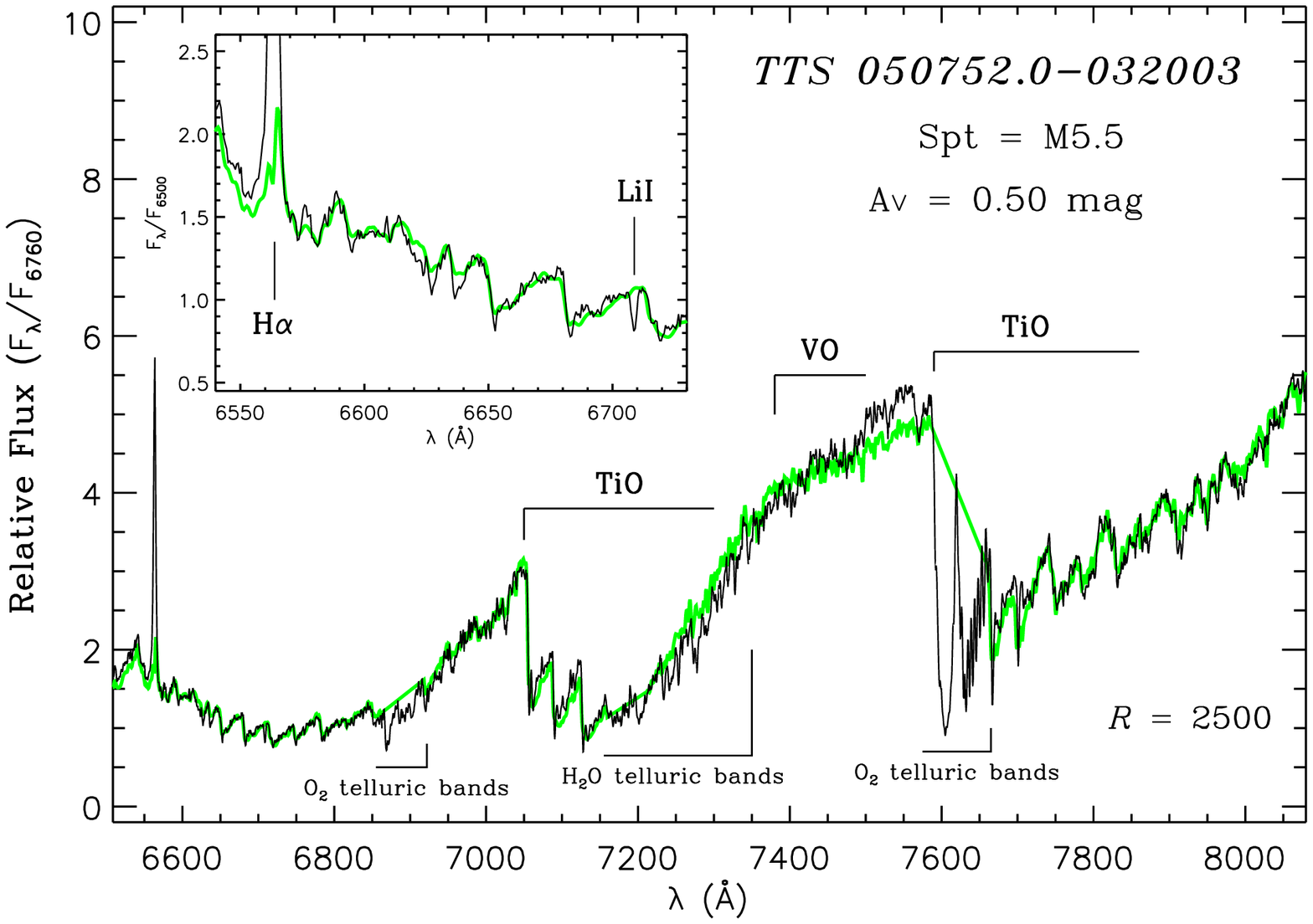}
 \caption{Low- (upper panel) and intermediate-resolution
          (lower panel) VIMOS spectra of TTS\,050752.0$-$032003. The
          observed spectra are displayed with thin lines,
          while the best-fitting M5.5 template, reddened with
          $A_{\mathrm V}=0.6$ and $0.5$ mag, respectively, is
          overplotted with a thick line. The spectra have
          been arbitrarily normalised to the flux at 6760~{\AA}.
          The O$_2$ absorption telluric bands near $\sim$6900 and
          $\sim$7600~{\AA}, as well as the H$_2$O one 
          near~$\sim$7250~{\AA} are marked. Note that these spectral
          regions have been excluded in the fitting procedure.
          The main TiO an VO bands are marked also. The Li\,{\sc i}
          $\lambda$6708~{\AA} absorption and the H$\alpha$
          emission lines are shown in more detail in the inset of
          the lower panel. Note the differences in H$\alpha$ and
          Li\,{\sc i} between the template and the observed spectra, 
          confirming the PMS nature of the star. Also note the
          good agreement between the two values of $A_{\mathrm V}$
          and the one derived from the SED fitting procedure
          (Table~\ref{Tab:Param}).}
 \label{VIMOS-spectra-fit}
\end{figure}

As an example, the FORS2 spectra of two PMS stars, namely RX\,J0507.6$-$0318  
and RX\,J0507.4$-$0320, as well as the low- and the intermediate-resolution
VIMOS spectra of TTS\,050752.0$-$032003 are shown in
Figure~\ref{FORS2-spectra-fit} and Figure~\ref{VIMOS-spectra-fit},
respectively; the best-fitting templates (thick lines)
are superimposed on the observed spectra (thin lines). Note that the
two VIMOS spectra yielded the same spectral type and a consistent
value of $A_{\mathrm V}$, well within the errors, regardless
of the different resolving power and wavelength coverage.

\clearpage

\clearpage

\section{SED fitting procedure}
\label{App-B}

The method we employed to determine both the reddening parameters (i.e.
$A_{\mathrm V}$ and $R_{\mathrm V}$) and the stellar intrinsic properties
(i.e. $R_{\star}$ and $L_{\star}$) of each PMS star, is based
on the use of appropriate stellar atmosphere models.
Taking into account the effective temperature range of the L1615/L1616
members (see Table~\ref{Tab:Param}), as well as the wide wavelength 
coverage required to encompass their SEDs
(see Table~\ref{Tab:abs-flux-calib}), we merged three distinct sets of
low-resolution synthetic spectra, adopting the following criteria:

\begin{itemize}

\item for the two intermediate-mass stars, namely the B9V-type
     HD\,293815 and the Herbig B3e-type star Kiso\,A-0974~15,
     we chose the grid of stellar atmosphere models computed by \citet{Kurucz79};

\item for sources covering the effective temperature interval between
     4000 and 10000 K we used the low-resolution synthetic spectra
     calculated by \citet{Hauschildt99} with their \emph{NextGen}
     model-atmosphere code;

\item for objects cooler than  4000~K we used the synthetic
     low-resolution \emph{StarDusty} spectra for low-mass stars and
     BDs by \citet{Allard00}. The \emph{StarDusty} spectra,
     which take into account dust grain formation in the model
     atmosphere due to an efficient gravitational settling process,
     are the most suitable for simulating cool objects, down
     to the very low-mass end. Indeed, \citet{Allard01} found that
     silicate dust grains can form abundantly in the outer
     atmospheric layers of the latest M dwarfs and BDs.

\end{itemize}

Each set of atmosphere models contains flux density spectra at the
stellar surface, computed for different value of gravity and
metallicity. We kept the gravity of the atmosphere models to a
fixed value of $\log(g)=4.0$  (cgs units), which is appropriate
for low-mass stars and BDs younger than $10$~Myr \citep{Chabrier00}.
Furthermore, this value closely matches the one expected for HAeBe
and early type main sequence stars \citep{Kurucz79,Schmidt-Kaler82}.
We also adopted a solar metallicity. 
However, we remind the reader that \citet{Romaniello02} have analysed the
influence of gravity and metallicity on the fitting-procedure and
found that neither of them affects the values of the derived
parameters significantly.

A grid of models matching the adopted effective temperature scale
(see Section~\ref{sec:Spec-Type-Class}) was first derived by
performing a linear interpolation of the merged set of synthetic
spectra.
In order to compare the observed SEDs with theoretical ones,
synthetic magnitudes in the same photometric systems in which the
observations were performed are required. Thus, for each adopted
temperature we derived a look-up table of synthetic $UBVR_CI_CJHKs$
absolute magnitudes by reddening the corresponding theoretical
spectrum with an increasing value of $A_{\mathrm V}$ (from 0 to 15
mag with a step of 0.05 mag) and $R_{\mathrm V}$ (from 1 to 9 with
a step of 0.1), and subsequently integrating the reddened spectrum
over each photometric pass-band. The extinction law by
\citet{Cardelli89} was adopted to perform this reddening procedure. Since
the model spectra are given as flux density at the star's surface,
the synthetic magnitudes were computed for a star\footnote{This choice
is only conventional, since at this step we were only interested in
the shape of the flux distribution.} with one solar radius, as seen
from a distance of $10$~pc. The Johnson-Cousins $UBVR_CI_C$ and
2MASS $JHKs$ transmission curves from {\it The Asiago Database on
Photometric Systems} \citep{Moro00, Fiorucci02}, as well as the
absolute flux calibration constants reported in
Table~\ref{Tab:abs-flux-calib} were used to integrate each
progressively reddened theoretical spectra.

Finally, the values of $A_{\mathrm V}$ and $R_{\mathrm V}$ of each
PMS star (see Table~\ref{Tab:Param}) were derived simultaneously by
fitting the corresponding observed SED to the look-up table of
reddened synthetic magnitudes having the same effective temperature
as the given object. In order to eliminate the magnitude shift
introduced by the unknown star's radius and distance, the observed
and the synthetic magnitudes were previously normalised to the
$I_C$-band. For PMS stars
showing clear evidence of near-infrared excess associated with
the presence of circumstellar material, the fit was performed to
the short-wavelength portion ($\lambda\leq\lambda_J$) of the SED,
which is less contaminated by near-infrared excess. The error on
$A_{\mathrm V}$ and $R_{\mathrm V}$ has been evaluated taking
into account both the accuracy on the spectral type classification
(Section~\ref{sec:Spec-Type-Class}) and the photometric errors on
the observed magnitudes.

\begin{figure}
 \centering
 \resizebox{11cm}{!}{\includegraphics[draft=false]{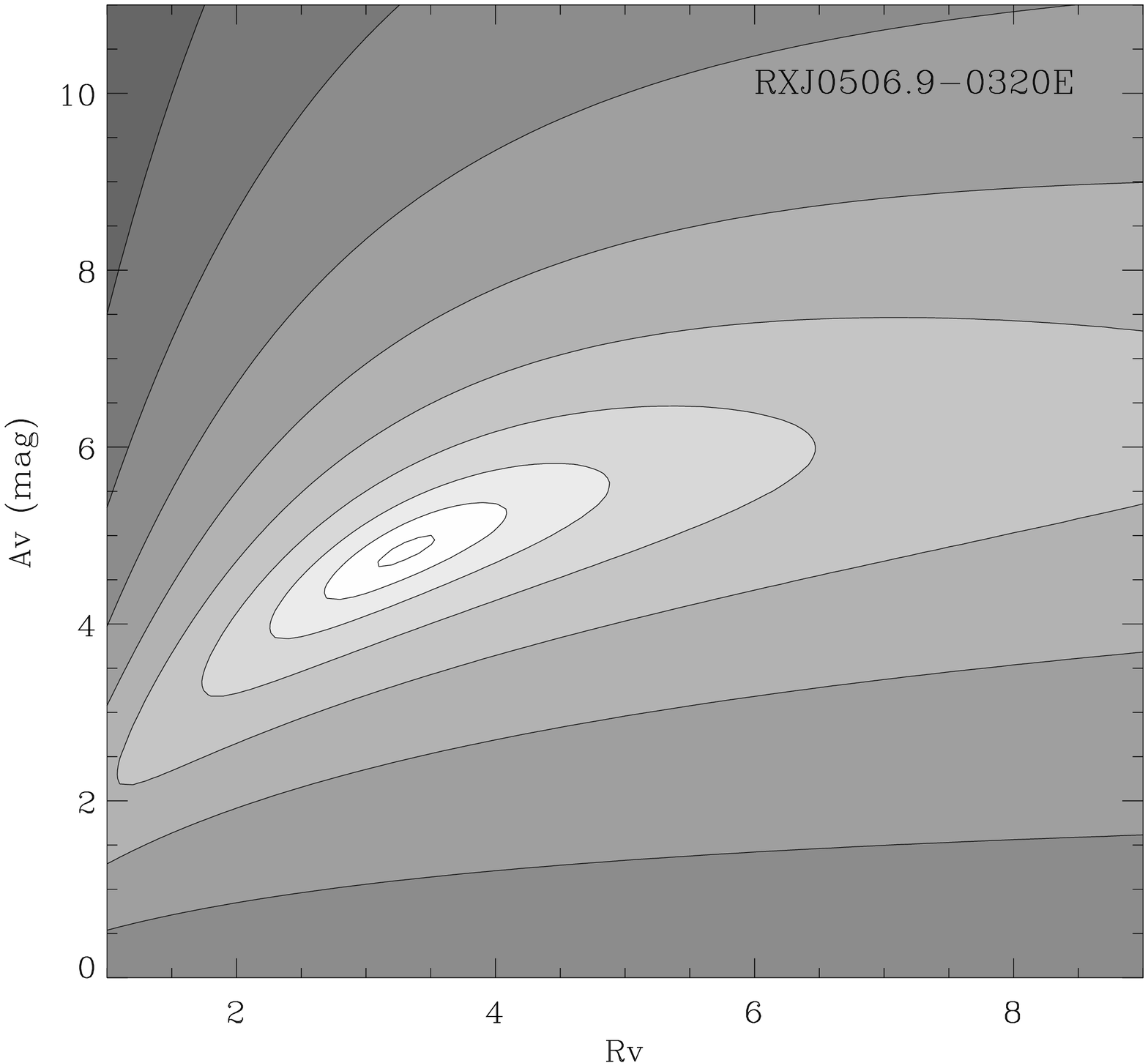}}
 \resizebox{11cm}{!}{\includegraphics[draft=false]{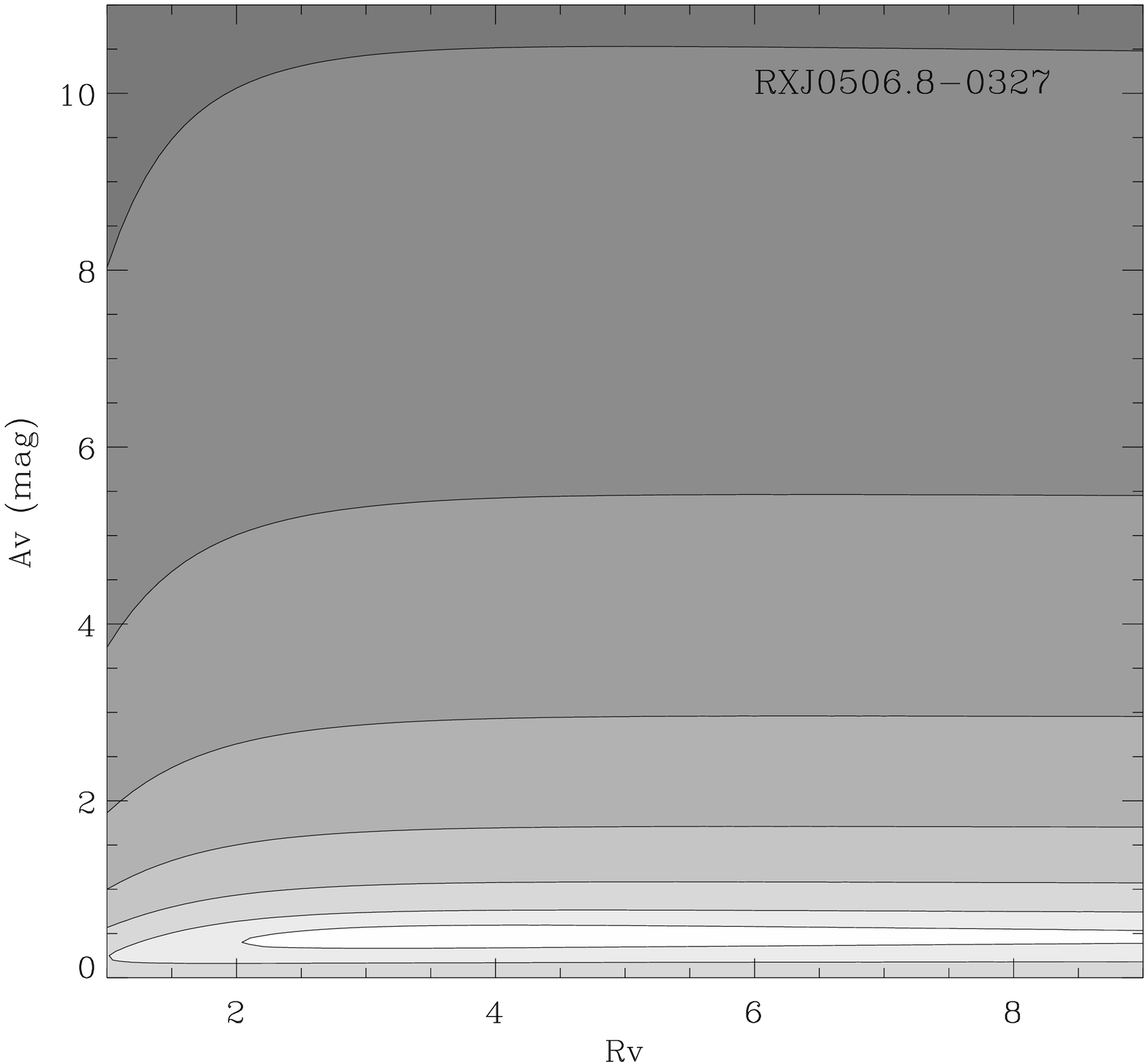}}
 \caption{$\chi^2$ contour plots produced by the SED fitting procedure for the
          two stars RX\,J0506.9$-$0320E and RX\,J0506.8$-$0327.}
 \label{Fig:CountourPlot}
\end{figure}

An example of the output of the fitting procedure is shown in the
upper panel of Figure~\ref{Fig:CountourPlot} , where a countour plot
of the $\chi^2$ value in the $R_{\mathrm V}$-$A_{\mathrm V}$ plane
is reported. As already stated in Section~\ref{sec:IntExt-Rad-Lum},
our two-parameter fitting procedure can be applied only to stars
which suffer a visual extinction larger than about 0.5~mag. For
low-reddened stars the value of $A_{\mathrm V}$ is fairly independent
of changes in $R_{\mathrm V}$.  As clearly evident in the lower panel
of Figure~\ref{Fig:CountourPlot}, in this case a unique solution could not be found
simultaneously in the $R_{\mathrm V}$-$A_{\mathrm V}$ plane.
Therefore, the value of $R_{\mathrm V}=3.1$, typical of the diffuse
interstellar medium, was assumed for all the object with
$A_{\mathrm V}\lesssim0.5$~mag.

Once the extinction parameters were determined, the value of
$R_{\star}$ for each PMS object was obtained from the magnitude
shifts required to match the observed magnitudes to the theoretical
ones, reddened with the best-fitting $A_{\mathrm V}$ and
$R_{\mathrm V}$. Only the photometric bands actually used in the
fitting procedure for each given object were employed to determine
its stellar radius. As suggested by \citet{Alcala04}, we made the
assumption that all the members of L1615/L1616 are located at
$450\pm20$ pc. Finally the value of $L_{\star}$ was derived assuming
a black body emission at the star's effective temperature and radius.
The associated uncertainties were derived taking into account the
errors on the distance and effective temperature.

\clearpage
\clearpage




{}

\clearpage

%%%%%%%%%%%%%%%%%%%%%%%%%%%%%%%%%%%%%%%%%   Table 1  %%%%%%%%%%%%%%%%%%%%%%%%%%%%%%%%%%%%%%%%%%%%%%%%%%%%%%%%%%%%%%%%%%%%

\begin{deluxetable}{cccccccccc}
  %\tabletypesize{\tiny}
  \tablecaption{Journal of the FORS2 and VIMOS spectroscopic observations.\label{Tab:Spec-Log}}
  \tablewidth{0pt}
  \rotate
  \tablehead{

  \colhead{Spectrograph}                 & \colhead{Period}            &
  \colhead{Grism}                        & \colhead{Order sorting}     &
  \colhead{$\lambda/\Delta\lambda^\ast$} & \colhead{$\lambda_{range}$} &
  \colhead{$T_{exp}$}                    & \colhead{$N_{mask}$}        &
  \colhead{$N_{object}$}                 \\

                                         &                             &
                                         & \colhead{filter}            &
                                         & \colhead{(\AA)}             &
  \colhead{(sec)}                        &                             &
  \colhead{per mask}

  }

  \startdata
         FORS2@VLT   &   Feb--Mar 2003      &  GRIS-1200R   &   GG435  &  2140  & 5700--7300$^\dag$ &    960      &     5       &     18--19    \\
         VIMOS@VLT   & Dec 2004--Mar 2005   &      HR\_red  &   GG475  &  2500  & 6300--8700$^\dag$ &   2000      &     3       &    $\sim80$   \\
         VIMOS@VLT   & Dec 2004--Mar 2005   &        MR     &   GG475  &   580  & 5000--8000~       &    900      &     6       &    $\sim120$  \\
  \enddata

  \tablenotetext{\ast}{Achieved at $\lambda=6500$~{\AA} with a $1{\arcsec}$ slit width.}
  \tablenotetext{\dagger}{Corresponding to a slit projected onto the centre of the detector.}

\end{deluxetable}

%%%%%%%%%%%%%%%%%%%%%%%%%%%%%%%%%%%%%%%%%%%%%%%%%%%%%%%%%%%%%%%%%%%%%%%%%%%%%%%%%%%%%%%%%%%%%%%%%%%%%%%%%%%%%%%%%%%%%%%%%

%%%%%%%%%%%%%%%%%%%%%%%%%%%%%%%%%%%%%%%%%   Table 2  %%%%%%%%%%%%%%%%%%%%%%%%%%%%%%%%%%%%%%%%%%%%%%%%%%%%%%%%%%%%%%%%%%%%

\begin{deluxetable}{lcccc}
  \tabletypesize{\footnotesize}
  \tablecaption{PMS stars in L1615/L1616. Coordinates are from 
                the 2MASS catalogue. The PMS objects identified 
                in this work and the previously known ones are 
                labelled with ``NEW'' and ``P.K.", respectively.
                For each star, additional designations found in 
                literature are listed in the last column.
                \label{Tab:List-PMS}}
  \tablewidth{0pt}

  \tablehead{

    \colhead{Star}                    & \colhead{$\alpha$(2000)}&
    \colhead{$\delta$(2000)}          & \colhead{Note}          &
    \colhead{Other Designations}                                  \\

                                      &   \colhead{(hh:mm:ss)}  &
    \colhead{(dd:mm:ss)}              &                         &

  }

\startdata
   1RXS~J045912.4$-$033711  &  04:59:14.59  &  $-$03:37:06.3  &  P.K.  &                                            \\
   1RXS~J050416.9$-$021426  &  05:04:15.93  &  $-$02:14:50.5  &  P.K.  &                                            \\
    TTS\,050513.5$-$034248  &  05:05:13.47  &  $-$03:42:47.8  &  NEW   &                                            \\
    TTS\,050538.9$-$032626  &  05:05:38.85  &  $-$03:26:26.4  &  NEW   &                                            \\
        RX\,J0506.6$-$0337  &  05:06:34.95  &  $-$03:37:15.9  &  P.K.  &                                            \\
    TTS\,050644.4$-$032913  &  05:06:44.42  &  $-$03:29:12.8  &  NEW   &                                            \\
    TTS\,050646.1$-$031922  &  05:06:46.05  &  $-$03:19:22.4  &  P.K.  &                                            \\
        RX\,J0506.8$-$0318  &  05:06:46.64  &  $-$03:18:05.6  &  P.K.  &                                            \\
\enddata
 
   \tablenotetext{a}{X-ray source by \citet{Alcala04}}
   \tablenotetext{b}{Mid-infrared source by \citet{Stanke02}}

\end{deluxetable}

%%%%%%%%%%%%%%%%%%%%%%%%%%%%%%%%%%%%%%%%%%%%%%%%%%%%%%%%%%%%%%%%%%%%%%%%%%%%%%%%%%%%%%%%%%%%%%%%%%%%%%%%%%%%%%%%%%%%%%%%%

%%%%%%%%%%%%%%%%%%%%%%%%%%%%%%%%%%%%%%%%%   Table 3  %%%%%%%%%%%%%%%%%%%%%%%%%%%%%%%%%%%%%%%%%%%%%%%%%%%%%%%%%%%%%%%%%%%%

\begin{deluxetable}{lccccccccc}
  \tabletypesize{\scriptsize}

  \tablecaption{Optical ($UBVR_CI_C$) and near-infrared ($JHKs$) photometry of the PMS stars in L1615/L1616.
                Labels are as explained in the footnotes.\label{Tab:phot}}
  \tablewidth{0pt}
  \rotate
  \tablehead{

   \colhead{PMS Star}       &   \colhead{$U$}  &   \colhead{$B$}  &   \colhead{$V$}  &  \colhead{$R_C$} & \colhead{$I_C$}  &  \colhead{$J$}   &  \colhead{$H$}   &  \colhead{$Ks$}  & \colhead{Notes}

   }

  \startdata
   1RXS~J045912.4$-$033711  &  12.74$\pm$0.05  &  12.45$\pm$0.04  &  11.73$\pm$0.03  &         NO       &  10.72$\pm$0.06  &  10.07$\pm$0.02  &   9.62$\pm$0.02  &   9.47$\pm$0.02  &   a,b,c  \\
   1RXS~J050416.9$-$021426  &  14.84$\pm$0.09  &  13.99$\pm$0.08  &  12.96$\pm$0.02  &         NO       &  11.59$\pm$0.03  &  10.66$\pm$0.02  &  10.10$\pm$0.02  &   9.98$\pm$0.03  &   a,c,d  \\
    TTS\,050513.5$-$034248  &         NO       &         ND       &         ND       &  18.97$\pm$0.02  &  17.03$\pm$0.02  &  15.12$\pm$0.04  &  14.60$\pm$0.05  &  14.20$\pm$0.06  &     c    \\
    TTS\,050538.9$-$032626  &         NO       &  18.88$\pm$0.02  &  17.30$\pm$0.02  &  16.21$\pm$0.02  &  14.67$\pm$0.01  &  13.29$\pm$0.02  &  12.60$\pm$0.02  &  12.39$\pm$0.02  &     c    \\
        RX\,J0506.6$-$0337  &  13.12$\pm$0.04  &  12.93$\pm$0.02  &  12.23$\pm$0.01  &         NO       &  11.24$\pm$0.05  &  10.66$\pm$0.02  &  10.30$\pm$0.03  &  10.20$\pm$0.02  &   a,b,c  \\
    TTS\,050644.4$-$032913  &         NO       &  19.33$\pm$0.01  &  17.69$\pm$0.02  &  16.68$\pm$0.02  &  15.08$\pm$0.01  &  13.64$\pm$0.03  &  13.04$\pm$0.03  &  12.78$\pm$0.03  &     c    \\
    TTS\,050646.1$-$031922  &         NO       &  18.88$\pm$0.01  &  17.34$\pm$0.02  &  16.09$\pm$0.01  &  14.81$\pm$0.01  &  12.60$\pm$0.01  &  11.09$\pm$0.01  &  10.22$\pm$0.01  &          \\
        RX\,J0506.8$-$0318  &         NO       &  16.26$\pm$0.01  &  14.86$\pm$0.02  &  13.91$\pm$0.01  &  13.01$\pm$0.01  &  11.95$\pm$0.01  &  11.22$\pm$0.01  &  10.97$\pm$0.01  &          \\
\enddata

    \tablenotetext{S~~}{Saturated object.}
    \tablenotetext{ND~~}{Object not detected.}
    \tablenotetext{NO~~}{Object not observed.}
    \tablenotetext{a~~}{$UBV$ from observations carried out at the Catania Astrophysical Observatory (Section~\ref{sec:OptPhot}).}
    \tablenotetext{b~~}{$I_C$ from TASS Mark IV catalogue \citep{Droege06}.}
    \tablenotetext{c~~}{$JHKs$ from \emph{2MASS} point source catalogue \citep{Cutri03}.}
    \tablenotetext{d~~}{$I_C$ from \emph{DENIS} catalogue \citep{Epchtein97}.}
    \tablenotetext{e~~}{$U$ and $B$ from \citet{Lee68}; $V$,$R_C$, and $I_C$ from \citet{Taylor89}.}
    \tablenotetext{f~~}{$U$ from \citet{Mundt80}.}
    \tablenotetext{g~~}{$UBVR_CI_C$ from \citet{Vieira03}.}
    \tablenotetext{h~~}{Source blended with another star of similar brightness: the WFI $BVR_CI_C$ photometry corresponds to the combined light of both stars.}
    \tablenotetext{i~~}{Weighted mean optical photometry from \citet{Cieslinski97}.}
    \tablenotetext{l~~}{$BVR_CI_C$ from \citet{Alcala96}.}
    \tablenotetext{m~~}{$J$ from \emph{DENIS} catalogue \citep[][converted to the 2MASS photometric system using the transformation equations provided by \citet{Carpenter01}]{Epchtein97}.}
    \tablenotetext{n~~}{$HKs$ from \emph{2MASS} point source catalogue \citep{Cutri03}.}
    \tablenotetext{p~~}{$UBV$ from \citet{Frasca03}.}

\end{deluxetable}

%%%%%%%%%%%%%%%%%%%%%%%%%%%%%%%%%%%%%%%%%%%%%%%%%%%%%%%%%%%%%%%%%%%%%%%%%%%%%%%%%%%%%%%%%%%%%%%%%%%%%%%%%%%%%%%%%%%%%%%%%

%%%%%%%%%%%%%%%%%%%%%%%%%%%%%%%%%%%%%%%%%   Table 4  %%%%%%%%%%%%%%%%%%%%%%%%%%%%%%%%%%%%%%%%%%%%%%%%%%%%%%%%%%%%%%%%%%%%

\begin{deluxetable}{lrrrccc}
 \tabletypesize{\footnotesize}

 \tablecaption{Spectral type, H$\alpha$ and Li\,{\sc i} equivalent widths of the PMS stars in L1615/L1616.
               Additional emission lines observed in the spectra are also tabulated.\label{Tab:SpT-W}}

 \tablewidth{0pt}

 \tablehead{
    \colhead{Star}          &   \colhead{~SpT}  &    \colhead{W(H$\alpha$)~~} &    \colhead{W(Li)~}    &   \colhead{Type}  &                   \colhead{Other emission lines}                   &  \colhead{Notes} \\
                            &                   &       \colhead{(\AA)~~~~}   &   \colhead{(m\AA)~}    &                   &                                                                    &
}
 \startdata
   1RXS~J045912.4$-$033711  &    G8   & $   1.80\pm0.20$  &  $355\pm10$  &   WTTS  &                                                          &  a  \\
   1RXS~J050416.9$-$021426  &    K3   & $   0.03\pm0.20$  &  $475\pm10$  &   WTTS  &                                                          &  a  \\
    TTS\,050513.5$-$034248  &  M5.5   & $  -9.80\pm0.40$  & $800\pm100$  &   WTTS  &                                                          &     \\
    TTS\,050538.9$-$032626  &  M3.5   & $  -4.90\pm0.20$  &  $350\pm20$  &   WTTS  &                        He\,{\sc i}                       &     \\
        RX\,J0506.6$-$0337  &    G9   & $   1.00\pm0.20$  &  $299\pm10$  &   WTTS  &                                                          &  a  \\
    TTS\,050644.4$-$032913  &  M4.5   & $  -4.70\pm0.30$  &  $560\pm40$  &   WTTS  &                                                          &     \\
    TTS\,050646.1$-$031922  &    K4   & $ -41.00\pm2.00$  &  $560\pm20$  &   CTTS  &          He\,{\sc i}, [N\,{\sc ii}], [S\,{\sc ii}]       &     \\
        RX\,J0506.8$-$0318  &  K8.5   & $  -3.80\pm0.20$  &  $570\pm30$  &   WTTS  &                        He\,{\sc i}                       &     \\
\enddata

    \tablenotetext{a}{W(H$\alpha$) and W(Li) from \citet{Alcala04}.}
    \tablenotetext{b}{Veiled CTTS.}
    \tablenotetext{c}{Spectral type from \citet{Sharpless52}.}
    \tablenotetext{e}{Spectral type from \citet{Vieira03}.}

\end{deluxetable}

%%%%%%%%%%%%%%%%%%%%%%%%%%%%%%%%%%%%%%%%%%%%%%%%%%%%%%%%%%%%%%%%%%%%%%%%%%%%%%%%%%%%%%%%%%%%%%%%%%%%%%%%%%%%%%%%%%%%%%%%%

%%%%%%%%%%%%%%%%%%%%%%%%%%%%%%%%%%%%%%%%%   Table 5  %%%%%%%%%%%%%%%%%%%%%%%%%%%%%%%%%%%%%%%%%%%%%%%%%%%%%%%%%%%%%%%%%%%%

\begin{deluxetable}{lccccr}
 \tabletypesize{\footnotesize}

 \tablecaption{Stellar parameters of the PMS stars in L1615/L1616.\label{Tab:Param}}

 \tablewidth{0pt}

 \tablehead{

    \colhead{PMS Star}      & \colhead{$\log T_{\mathrm eff}$} &      \colhead{$R_{\mathrm V}$}     & \colhead{$A_{\mathrm V}$} & \colhead{$R_{\star}$}   & \colhead{$\log L/L_{\odot}$} \\
                            &                                  &                                    &       \colhead{(mag)}     & \colhead{($R_{\odot}$)} &
 }

 \startdata
   1RXS~J045912.4$-$033711  &          $3.746\pm0.009$         &  $3.1\pm0.1$\tablenotemark{\star}  &       $0.30\pm0.12$       &      $2.63\pm0.12$      &      $ 0.78\pm0.05$  \\
   1RXS~J050416.9$-$021426  &          $3.675\pm0.014$         &  $3.1\pm0.1$\tablenotemark{\star}  &       $0.35\pm0.22$       &      $2.35\pm0.11$      &      $ 0.40\pm0.07$  \\
    TTS\,050513.5$-$034248  &          $3.494\pm0.010$         &  $3.1\pm0.1$\tablenotemark{\star}  &       $0.49\pm0.20$       &      $0.55\pm0.03$      &      $-1.60\pm0.06$  \\
    TTS\,050538.9$-$032626  &          $3.531\pm0.009$         &  $3.1\pm0.1$\tablenotemark{\star}  &       $0.05\pm0.10$       &      $1.08\pm0.05$      &      $-0.85\pm0.05$  \\
        RX\,J0506.6$-$0337  &          $3.733\pm0.013$         &  $3.1\pm0.1$\tablenotemark{\star}  &       $0.05\pm0.10$       &      $1.92\pm0.09$      &      $ 0.45\pm0.06$  \\
    TTS\,050644.4$-$032913  &          $3.513\pm0.009$         &  $3.1\pm0.1$\tablenotemark{\star}  &       $0.00 + 0.05$       &      $1.00\pm0.05$      &      $-0.99\pm0.05$  \\
    TTS\,050646.1$-$031922  &          $3.662\pm0.026$         &  $4.2\pm0.4$~                      &       $4.33\pm0.35$       &      $1.91\pm0.11$      &      $ 0.17\pm0.11$  \\
        RX\,J0506.8$-$0318  &          $3.597\pm0.008$         &  $3.1\pm0.1$\tablenotemark{\star}  &       $0.00 + 0.05$       &      $1.62\pm0.07$      &      $-0.24\pm0.05$  \\
\enddata

\tablenotetext{\star}{Assumed (see Section~\ref{sec:IntExt-Rad-Lum}).}

\end{deluxetable}

%%%%%%%%%%%%%%%%%%%%%%%%%%%%%%%%%%%%%%%%%%%%%%%%%%%%%%%%%%%%%%%%%%%%%%%%%%%%%%%%%%%%%%%%%%%%%%%%%%%%%%%%%%%%%%%%%%%%%%%%%

%%%%%%%%%%%%%%%%%%%%%%%%%%%%%%%%%%%%%%%%%   Table 6  %%%%%%%%%%%%%%%%%%%%%%%%%%%%%%%%%%%%%%%%%%%%%%%%%%%%%%%%%%%%%%%%%%%%

\begin{table}
  \caption{Effective wavelengths and absolute flux calibration constants for the Joh\-nson-Cousins and 2MASS
           passbands. The main reference papers are listed in the last column.}
  \begin{center}
  \begin{tabular}{cccc}
  \hline
  \hline
  \noalign{\medskip}
         Filter         & $\lambda_{\mathrm eff}$ &      $F^0_\lambda$      &     Reference     \\
                        &      ($\mu$m)           & ($erg/s^{2}/cm^{2}/$\AA)&                   \\
  \noalign{\medskip}
  \hline
  \noalign{\smallskip}
         $U_J$          &       0.36          &   4.350$\cdot 10^{-9} $ & \citet{Johnson65} \\
         $B_J$          &       0.44          &   7.200$\cdot 10^{-9} $ & \citet{Johnson65} \\
         $V_J$          &       0.55          &   3.920$\cdot 10^{-9} $ & \citet{Cousins76} \\
         $R_C$          &       0.64          &   2.254$\cdot 10^{-9} $ & \citet{Cousins76} \\
         $I_C$          &       0.79          &   1.196$\cdot 10^{-9} $ & \citet{Cousins76} \\
          $J$           &       1.24          &   3.129$\cdot 10^{-10}$ & \citet{Cohen03}   \\
          $H$           &       1.66          &   1.133$\cdot 10^{-10}$ & \citet{Cohen03}   \\
         $Ks$           &       2.16          &   4.283$\cdot 10^{-11}$ & \citet{Cohen03}   \\
  \noalign{\smallskip}
  \hline
  \end{tabular}
  \end{center}
  \label{Tab:abs-flux-calib}
\end{table}

\clearpage

%%%%%%%%%%%%%%%%%%%%%%%%%%%%%%%%%%%%%%%%%%%%%%%%%%%%%%%%%%%%%%%%%%%%%%%%%%%%%%%%%%%%%%%%%%%%%%%%%%%%%%%%%%%%%%%%%%%%%%%%%

%%%%%%%%%%%%%%%%%%%%%%%%%%%%%%%%%%%%%%%%%   Table 7  %%%%%%%%%%%%%%%%%%%%%%%%%%%%%%%%%%%%%%%%%%%%%%%%%%%%%%%%%%%%%%%%%%%%

\begin{deluxetable}{lcccccc}
 \tabletypesize{\footnotesize}

 \tablecaption{Masses and ages for the PMS stars in L1615/L1616 as inferred by the evolutionary models by \citet{Baraffe98}
               \& \citet{Chabrier00}, \citet{Dantona97} and \citet{Palla99}.\label{Tab:mass_age}}

 \tablewidth{0pt}

 \tablehead{

     \colhead{PMS Star}        & \colhead{$M_{Ba98+Ch00}$} & \colhead{$Age_{Ba98+Ch00}$} &  \colhead{$M_{DM97}$}   & \colhead{$Age_{DM97}$} &  \colhead{$M_{PS99}$}   & \colhead{$Age_{PS99}$} \\

                               & \colhead{($M_{\odot}$)}   &       \colhead{(Myr)}       & \colhead{($M_{\odot}$)} &     \colhead{(Myr)}    & \colhead{($M_{\odot}$)} &    \colhead{(Myr)}
 }

 \startdata
      1RXS~J045912.4$-$033711  &          $>1.40$          &              $-$            &           2.00          &          5.00          &          1.70           &           3.50         \\
      1RXS~J050416.9$-$021426  &          $>1.40$          &              $-$            &           0.95          &          0.70          &          1.60           &           3.00         \\
       TTS\,050513.5$-$034248  &            0.12           &             6.00            &           0.17          &         10.00          &          0.14           &          10.00         \\
       TTS\,050538.9$-$032626  &            0.35           &             4.04            &           0.25          &          2.20          &          0.26           &           2.80         \\
           RX\,J0506.6$-$0337  &            1.35           &            14.00            &           1.65          &          5.50          &          1.40           &           8.00         \\
       TTS\,050644.4$-$032913  &            0.25           &             3.17            &           0.20          &          2.50          &          0.19           &           2.70         \\
       TTS\,050646.1$-$031922  &          $>1.40$          &              $-$            &           0.75          &          1.00          &          1.35           &           4.00         \\
           RX\,J0506.8$-$0318  &            1.05           &             5.00            &           0.50          &          1.00          &          0.68           &           3.00         \\
\enddata

\end{deluxetable}

\clearpage

%%%%%%%%%%%%%%%%%%%%%%%%%%%%%%%%%%%%%%%%%%%%%%%%%%%%%%%%%%%%%%%%%%%%%%%%%%%%%%%%%%%%%%%%%%%%%%%%%%%%%%%%%%%%%%%%%%%%%%%%%

%%%%%%%%%%%%%%%%%%%%%%%%%%%%%%%%%%%%%%%%%   Table 8  %%%%%%%%%%%%%%%%%%%%%%%%%%%%%%%%%%%%%%%%%%%%%%%%%%%%%%%%%%%%%%%%%%%%

\begin{deluxetable}{ccccc}

    \tablecaption{ ~$\alpha$ slopes of the IMF in different clusters and associations.\label{Tab:IMF}}
    \tablewidth{0pt}

    \tablehead{
          \colhead{Cluster} &     \colhead{Age}     &    \colhead{Mass range}  &     \colhead{$\alpha$}   &     \colhead{Ref.}   \\
                            &    \colhead{(Myr)}    &  \colhead{($M_{\odot}$)} &                          &
    }

\startdata

            M35   &      150    &     0.1-0.4    &     0.58       &     a      \\
    \noalign{\smallskip}
    $\alpha$ Per  &       80    &    0.06-0.2    &     0.56       &     b      \\
    \noalign{\smallskip}
      Pleiades    &      120    &    0.03-0.45   & 0.60$\pm$0.11  &     c      \\
    \noalign{\smallskip}
 $\lambda$ Orionis&        5    &    0.02-1.2    & 0.60$\pm$0.06  &     d      \\
    \noalign{\smallskip}
 $\sigma$ Orionis &        5    &    0.013-0.2   & 0.8$\pm$0.4    &     e      \\
    \noalign{\smallskip}
     Trapezium    &      0.4    &    0.035-0.56  &    $\sim$0.3   &     f      \\
                  &             &     0.25-3     &    $\sim$0.7   &            \\
    \noalign{\smallskip}
    $\rho$ Oph    &    0.1-1    &     0.02-0.4   &    $\sim$0.5   &     g      \\
\enddata
\tablerefs{
a) \citet{Barrado98b}; b) \citet{Barrado02b}; c) \citet{Moraux03}; d) \citet{Barrado04b}; e) \citet{Bejar01};
f) \citet{Luhman00}; g) \citet{Luhman99b}; h) \citet{Tej02}; i) \citet{Briceno02}; j) \citet{Lopez04}; k) \citet{Comeron00};
l) \citet{Spezzi08}.}
\end{deluxetable}

\end{document}